%% file: gykim.tex
\documentclass[sigconf,9pt]{acmart}
\usepackage[english]{babel}

\setcopyright{acmcopyright}
\acmYear{2023}\copyrightyear{2023}
\acmConference[ACM SIGCOMM '23]{ACM SIGCOMM 2023 Conference}{September 10--14, 2023}{New York, NY, USA}
\acmBooktitle{ACM SIGCOMM 2023 Conference (ACM SIGCOMM '23), September 10--14, 2023, New York, NY, USA}
\acmPrice{15.00}
\acmDOI{10.1145/3603269.3604820}
\acmISBN{979-8-4007-0236-5/23/09}
\usepackage{epsfig,subfig,graphicx,endnotes,multirow,amsmath,algorithm,algpseudocode}
\usepackage{xcolor,xspace}
\usepackage{url}
\newcommand\StateX{\Statex\hspace{\algorithmicindent}}%
\usepackage{balance}
\usepackage{enumitem}
\usepackage{flushend}
\usepackage[T1]{fontenc}
\usepackage{color,soul}
\usepackage{amsthm}

\newcommand{\sys}{NetClone\xspace}
\newcommand{\sota}{LÆDGE\xspace}
\newcommand{\cc}{C-Clone\xspace}
\newcommand{\etal}{\textit{et al.}\xspace}

\begin{document}
\title{\sys: Fast, Scalable, and Dynamic Request Cloning for Microsecond-Scale RPCs}
\author{Gyuyeong Kim}
 \affiliation{%
   \institution{Sungshin Women's University}
  \country{South Korea}
 }
 \email{gykim@sungshin.ac.kr}

\renewcommand{\shortauthors}{Gyuyeong Kim}

\begin{abstract}
\input{./sections/abstract.tex} 
\end{abstract}

\maketitle

\input{./sections/introduction.tex} 
\input{./sections/motivation.tex}

\input{./sections/design.tex} 
\input{./sections/implementation.tex} 
\input{./sections/evaluation.tex}

\input{./sections/relatedwork.tex}

\input{./sections/conclusion.tex}  
\input{./sections/ack.tex}
\bibliographystyle{ACM-Reference-Format}
\bibliography{./gykim}
\end{document}

%% file: sections/abstract.tex
Spawning duplicate requests, called cloning, is a powerful technique to reduce tail latency by masking service-time variability.
However, traditional client-based cloning is static and harmful to performance under high load, while a recent coordinator-based approach is slow and not scalable.
Both approaches are insufficient to serve modern microsecond-scale Remote Procedure Calls (RPCs).
To this end, we present \sys, a request cloning system that performs cloning decisions dynamically within nanoseconds at scale.
Rather than the client or the coordinator, \sys performs request cloning in the network switch by leveraging the capability of programmable switch ASICs.
Specifically, \sys replicates requests based on server states and blocks redundant responses using request fingerprints in the switch data plane.
To realize the idea while satisfying the strict hardware constraints, we address several technical challenges when designing a custom switch data plane.
\sys can be integrated with emerging in-network request schedulers like RackSched.
We implement a \sys prototype with an Intel Tofino switch and a cluster of commodity servers.
Our experimental results show that \sys can improve the tail latency of microsecond-scale RPCs for synthetic and real-world application workloads and is robust to various system conditions. 

%% file: sections/introduction.tex
\section{Introduction \label{introduction}}
Today's online services are made up of multiple microservices that communicate with each other using Remote Procedure Calls (RPCs), allowing access to functions and data as if they were local~\cite{alto,breakwater}.
These services often have strict Service Level Objectives (SLOs) that require underlying data center systems to provide high throughput with \textit{microsecond-scale} tail latency~\cite{attackofus,dean13,breakwater,shinjuku,mcclure22}.
This is because RPCs are getting smaller, and their runtime is generally an order of microseconds~\cite{shinjuku,alto,erpc}. 
Unfortunately, RPC requests often experience excessive tail latency even if the request is the same~\cite{dean13,laedge}.
One of the causes is unexpected variability in service times, which stems from various factors (e.g., load fluctuation, background tasks, interference among applications, and garbage collection~\cite{powerofd,dolly,rpcvalet,laedge}).

Request cloning is a powerful technique to mask service-time variability.
The traditional client-based cloning always sends redundant requests (typically 2~\cite{laedge,powerofd}) to multiple servers and only accepts the faster response.
Owing to its simplicity and efficiency, the cloning technique has been employed in various domains~\cite{vulimiri13, dolly, xu14a,laedge,powerofd,redundancy}.
One limitation is that it does not always result in improved performance.
The latency is improved only within a sweet spot, and the system performance is rather degraded beyond a certain threshold load~\cite{laedge,vulimiri13}.
This is not surprising because redundant requests add extra load to servers.
Redundancy also doubles the packet processing overhead for clients, reducing the performance gain~\cite{vulimiri13}.

A recent solution~\cite{laedge} addresses the limitation by using a centralized coordinator, which dynamically clones requests only if at least two servers are idle.
Thanks to dynamic cloning, the performance is not degraded under high load.
However, it is not enough to serve microsecond-scale workloads.
This is because the coordinator incurs microseconds of additional latency overhead.
It is also hard to scale out as throughput grows because of the limited capability of the coordinator CPU.
Therefore, its target workload is millisecond-scale workloads with limited throughput. 
In this context, we ask the following question: \textit{how can we perform dynamic request cloning quickly at scale for microsecond-scale RPCs?}

As the answer to the question, we present \sys, a new request cloning system for microsecond-scale RPCs.
To serve these workloads with high throughput and low tail latency at the cluster-level, cloning decisions should be made in a nanosecond-scale with scalability.
However, achieving this in software is difficult because this is beyond the capability of modern CPUs even with advanced networking like RDMA.
For this reason, \sys performs request cloning in hardware.
Specifically, we dynamically clone requests and filter redundant slower responses in the Top-of-Rack (ToR) switch by leveraging the capability of programmable switch ASICs like Intel Tofino~\cite{tofinonodate}.
The switch can process a few billion packets per second, and it takes only hundreds of nanoseconds to process a single packet.
Therefore, with dynamic in-network request cloning, we can avoid latency overhead and a potential performance bottleneck, which are caused by the cloning coordinator.

However, transforming the high-level idea into a working system is not straightforward because of various technical challenges as follows.
First, we need to know server states (i.e., busy or idle) for cloning decisions.
While the existing solution~\cite{laedge} can guarantee the idleness of servers by queueing requests in the coordinator, we cannot directly implement it in the programmable switch because of limited memory space.
To address this, we make response packets piggyback the state of the server by lookup the vacancy of the request queue, and the switch stores the state in the switch memory.
The switch replicates requests only if two candidate servers are idle.
Unfortunately, the actual server state may be different due to the time gap.
Therefore, we design a server-side mechanism that drops the cloned request if the actual state is busy.

Second, we need to access the server state table twice to get the state of the candidate servers.
However, this is not possible with the current programmable switch ASIC that makes packets go through processing stages sequentially.
In particular, it requires two stages to access the state table twice, but the table is statically allocated in the first stage.
To overcome this limitation, we put a shadow table in the second stage, a copy of the state table.
Similarly, we cannot assign the destination IP to the cloned request at the time of cloning.
To address this, we recirculate the cloned request by forwarding the clone to a port in loopback mode.

Lastly, we need to block the slower response because it reduces the performance gain by causing unnecessary packet processing in the client.
The challenge here is that memory footprint and hash collisions should be minimized.
To address this, we make a filter table using the hash index, which can be reused by multiple requests.
For the faster response of a request, the switch puts its request ID in the filter table as a fingerprint.
In contrast, the switch drops the slower response of the request if the table slot contains the same request ID, since the switch knows that the faster response is already processed.
To handle hash collisions, we use multiple filter tables with randomized table indices for requests.

\sys is in line with emerging in-network computing solutions~\cite{netlr,racksched,jin18,yu20,harmonia}.
We believe that a tier of coordinators like load balancers between clients and servers should be integrated into the network switch because we can eliminate performance overhead and save costs of hardware and software required to build and maintain coordinator nodes as well.
In this context, \sys is a further advance to realize the vision of in-network computing, not just another case to show the benefit of in-network acceleration.
To demonstrate this, we show that \sys can be integrated with RackSched~\cite{racksched}, a recent in-network request scheduler.

We implement a prototype of \sys on an Intel Tofino switch.
\sys consumes 4.77\% of the switch memory because we store small soft states in switch memory, which are generally server state information and request IDs in the filter table.
To evaluate \sys, we build a testbed consisting of 8 commodity servers and a 6.5Tbps Intel Tofino programmable switch.
We conduct a series of extensive experiments with a combination of synthetic\, Redis~\cite{redis} and Memcached~\cite{memcached} workloads.
Our key findings include: 1) \sys can provide lower tail latency compared to the baseline, and has higher throughput than the client-based cloning and \sota~\cite{laedge}, the state-of-the-art coordinator-based cloning solution; 2) \sys can make synergy with RackSched~\cite{racksched} for various workload conditions; 3) \sys is robust to system conditions.

In summary, this work makes the following contributions.
\begin{itemize}[noitemsep]
\item{
We propose \sys, a request cloning system that provides dynamic, scalable, and fast request cloning and redundant response filtering to reduce the tail latency for modern microsecond-scale RPC workloads.
\sys shows that programmable switches are a vantage point that can be used to accelerate applications with microsecond-scale latencies.
}
\item{
We address various technical challenges to design a custom switch data plane that clones requests, filters redundant responses, and tracks server states within the strict hardware constraints of switch ASICs.
}
\item{
We implement a \sys prototype with a commodity programmable switch and conduct a series of extensive testbed experiments to demonstrate the efficiency and robustness of \sys.
}
\end{itemize}

The remainder of the paper is organized as follows.
In Section~\ref{motivation}, we describe the motivation of this work.
Section~\ref{design} provides the design of \sys.
We present implementation and evaluation results in Section~\ref{implementation} and Section~\ref{evaluation}, respectively.
We discuss related work in Section~\ref{relatedwork}.
Lastly, we conclude our work in Section~\ref{conclusion}.

%% file: sections/motivation.tex
\section{Background and Motivation \label{motivation}}
In this section, we provide background on request cloning and motivate the necessity of in-network request cloning for microsecond-scale RPCs.

\subsection{Latency Variability in RPCs}
Modern online services consist of a set of microservices, and the interaction between the microservice applications is often done by RPCs~\cite{servicefabric,breakwater}.
To guarantee good user experience, online services have strict SLOs, which are generally expressed as tail latency.
The runtime of RPCs is typically short as a few to tens of microseconds~\cite{shinjuku,alto}.
Therefore, data center systems that host the services are expected to provide low tail latency with high throughput.

Unfortunately, variability in service times makes it challenging to ensure low tail latency.
The processing latency of requests in a server is stochastic and sometimes can be 15 times larger than the median latency~\cite{laedge}.
Various factors contribute to the service-time variability, which include load fluctuation, interrupts, garbage collection, background tasks, OS scheduling, power management, and so on~\cite{laedge,dean13,vulimiri13, dolly, xu14a,rpcvalet,chronos}.
Therefore, the service time of RPCs typically follows a heavy-tailed distribution~\cite{rpcvalet,dean13,breakwater}, which may violate the SLO of the services.

\subsection{Cloning for Microsecond-Scale RPCs}
One efficient technique to mask service-time variability is request cloning.
The idea is simple as follows.
The client sends multiple copies of a request to different servers and takes the fastest response.
Optionally, the client may cancel unfinished slower requests.
Recent results show that two clones are enough, and canceling slower requests does not bring meaningful benefits~\cite{laedge}.
Owing to its simplicity and efficiency, cloning has been adopted in various works over the past decade~\cite{vulimiri13, dolly, xu14a,laedge,powerofd,redundancy}.
There are two approaches for cloning.
One is traditional client-based cloning~\cite{vulimiri13}, which we call \cc in short, and the other is coordinator-based cloning~\cite{laedge}.
Unfortunately, these approaches are not enough to serve microsecond-scale RPC workloads.

\begin{figure}[t!]
\centering\hfill
\subfloat[C-Clone~\cite{vulimiri13}]{\includegraphics[width=0.3\linewidth]{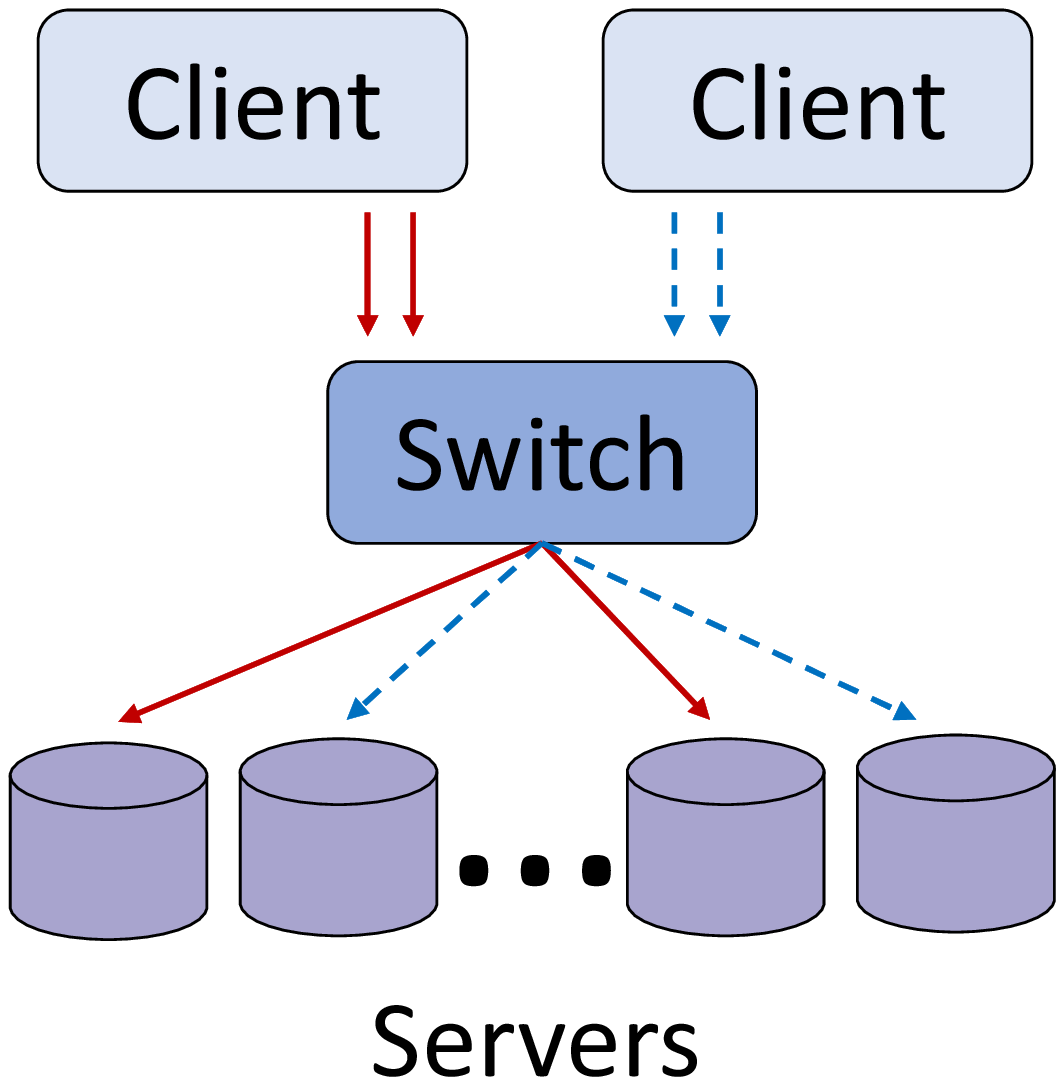}}\hfill
\subfloat[\sota~\cite{laedge}]{\includegraphics[width=0.3\linewidth]{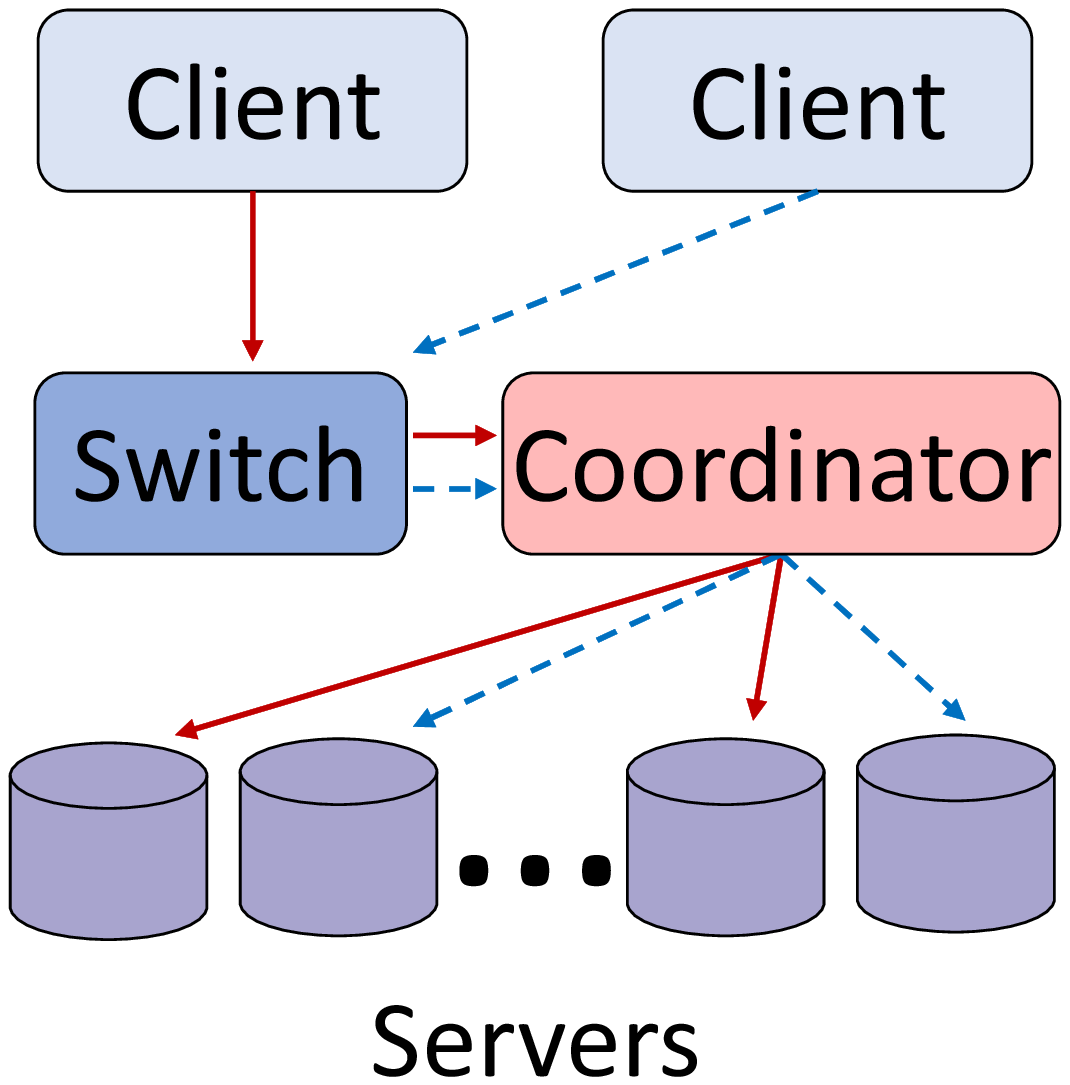}}\hfill
\subfloat[\sys]{\includegraphics[width=0.3\linewidth]{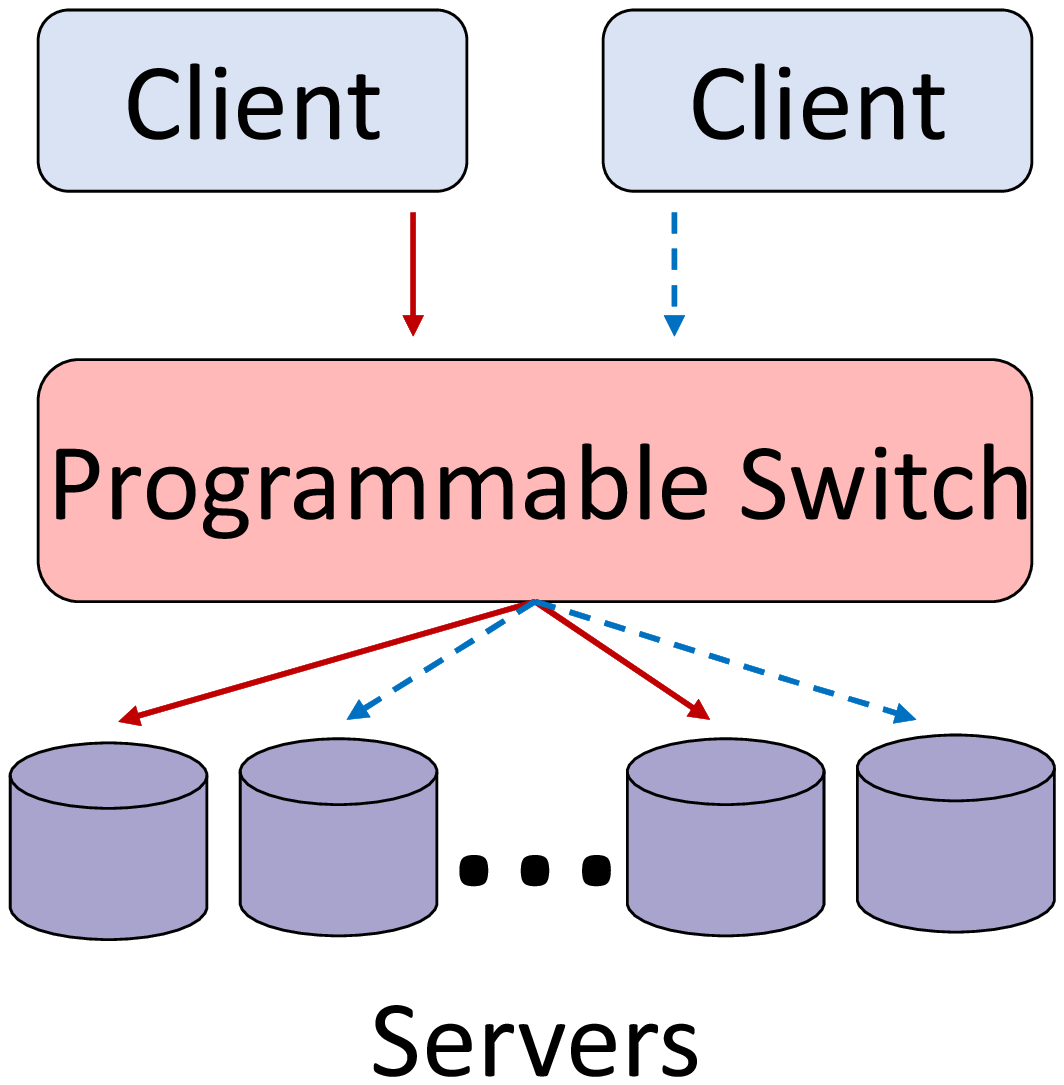}}\hfill
\caption{Different approaches for requeset cloning.
\label{fig:motiv-com}}
\end{figure}

\textbf{Client-based cloning (\cc).}
With this, clients perform request cloning in a distributed manner~\cite{vulimiri13} as illustrated in Figure~\ref{fig:motiv-com} (a).
The client typically sends two duplicate requests to servers.
One limitation is that cloning is only beneficial within a specific load range.
The latency is degraded significantly after a tipping point, which typically lies between 25\% and 50\% of the load.
This is due to the static and load-agnostic cloning of the client, as it always sends duplicate requests regardless of system load.
This static cloning also degrades maximum throughput by half as server loads become double.

\textbf{Coordinator-based cloning.}
This approach uses a coordinator node to perform request cloning in a centralized manner as shown in Figure~\ref{fig:motiv-com} (b).
\sota~\cite{laedge} is the state-of-the-art coordinator-based solution.
Unlike \cc, \sota is dynamic and load-aware.
The coordinator only replicates requests if at least two servers are idle.
If only one server is available, the request is forwarded without replication.
In the case where all servers are busy, the coordinator enqueues the request in a request queue and waits for an idle server.
The buffered request is dispatched to a server upon receiving a response.

Unfortunately, this is still far from a solution for microsecond-scale RPCs.
The cloning decision needs to be as fast as possible since RPCs want to be processed as if they are local functions.
However, it takes an order of microseconds to perform request cloning in the coordinator.
Because of this, \sota targets millisecond-scale workloads, which can tolerate the latency overhead.

The other limitation is that the coordinator is not scalable because it relies on the CPU to handle requests. 
Unfortunately, the CPU has inherently limited performance even with kernel-bypass networking like RDMA, which can reduce CPU usage for packet processing.
Therefore, the coordinator can be a performance bottleneck easily and provide limited throughput for only a few servers.
Furthermore, the \sota coordinator should process redundant slower responses to dispatch another request, making throughput worse.
It is possible to use multiple coordinators to scale out.
However, this causes burdensome costs to build and maintain a tier of coordinators.

\subsection{The Case for In-Network Cloning}
\textbf{Design goal and key idea.}
Our goal is to perform request cloning dynamically and quickly at scale for microsecond-scale RPCs.
The key idea to achieve the goal is to perform cloning decisions in the switch by leveraging the capability of programmable switch ASICs like Intel Tofino~\cite{tofinonodate} and Cavium Xpliant~\cite{xpliantnodate}.
We can achieve high performance using the switch since it is optimized for packet processing.
In particular, a switch can process a few billion packets per second, whereas a commodity server can handle a few million packets per second.
Furthermore, the per-packet processing delay is guaranteed in hundreds of nanoseconds.
Therefore, we propose \sys, an in-network dynamic request cloning system as shown Figure~\ref{fig:motiv-com} (c).

\textbf{Comparison to existing works.}
Table~\ref{table:comparison} summarizes the difference between \sys and the existing solutions.
\cc can scale out to multiple servers and does not incur excessive latency overhead for cloning decisions.
However, as it statically replicates requests regardless of system load, throughput is limited.
Despite dynamic cloning, \sota does not provide scalability, high throughput, and low latency overhead as it uses a server-based cloning coordinator.
Unlike the existing works, \sys can clone requests dynamically at scale with high throughput and a nanosecond-scale latency overhead as cloning is performed in the network switch.

\textbf{Challenges.}
Designing an in-network cloning system does not mean merely implementing the existing dynamic cloning mechanism on the network switch.
This is because the switch has strict resource constraints and timing requirements.
When designing a custom switch data plane, we should address several technical challenges as follows.
\begin{itemize}[noitemsep]
\item{
The switch contains only 10-20MB of limited memory.
This implies that we cannot queue requests in the switch memory as \sota does, and we need a new mechanism to check whether servers are idle or busy.
}
\item{
It is impossible to access data stored in the memory twice for a single pass because each data is statically allocated to a specific stage at compile time.
This means that it is challenging to check the state of two candidate servers for cloning decisions.
}
\item{
In a similar vein, we should carefully design a mechanism to filter redundant slower responses while minimizing memory footprints.
}
\end{itemize}

\begin{table}[t]  
\center
\caption{Comparison to existing works.\label{table:comparison}} 
\footnotesize
\begin{tabular}{|c|c|c|c|}
\hline
\textbf{ } & \textbf{C-Clone~\cite{vulimiri13}} & \textbf{\sota~\cite{laedge}}  & \textbf{\sys}\\ \hline\hline
Cloning point & Client & Coordinator & Switch  \\
Dynamic cloning & $\times$ & $\surd$ & $\surd$  \\
Scalability & $\surd$  & $\times$& $\surd$  \\
High throughput & $\times$ & $\times$& $\surd$  \\
Low latency overhead& $\surd$  & $\times$& $\surd$  \\
\hline
\end{tabular}
\end{table}

%% file: sections/design.tex
\section{\sys Design \label{design}}
\subsection{\sys Architecture}
\begin{figure}[t!]
\centering
\includegraphics[width=8.0cm]{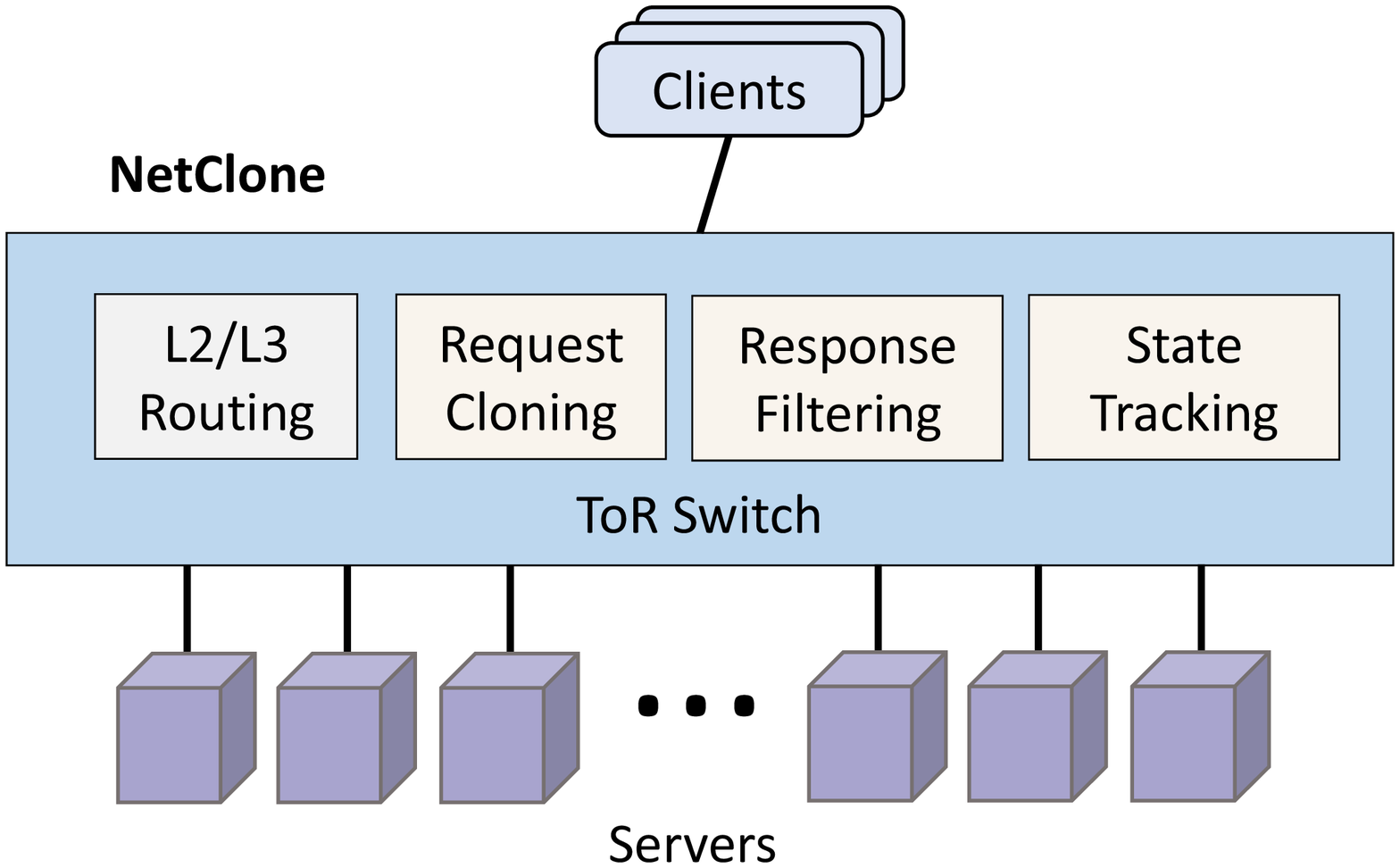}
\caption{\sys system architecture. \label{fig:overview}}
\end{figure}

As illustrated in Figure~\ref{fig:overview}, the \sys architecture consists of the switch data plane, clients, and servers.

\textbf{Switch data plane.}
The core of the \sys architecture is the switch data plane.
We design three custom modules, which are triggered only for \sys packets.
The request cloning module decides whether to replicate requests based on server states.
The response filtering module blocks redundant slower responses using request fingerprints.
The state tracking module updates the server states upon receiving responses to track the latest state information.
Meanwhile, our switch data plane can perform packet forwarding with the traditional L2/L3 routing module.

\textbf{Clients and servers.}
To support \sys, we need modifications on clients and servers to insert metadata (e.g., server state) into the \sys header, which resides between the L4 header and the application payload.
Note that integrating \sys with existing RPC frameworks needs careful investigation because it may cause interference between request cloning and existing functionality in the framework.

\begin{figure}[t!]
\centering
\includegraphics[width=8.0cm]{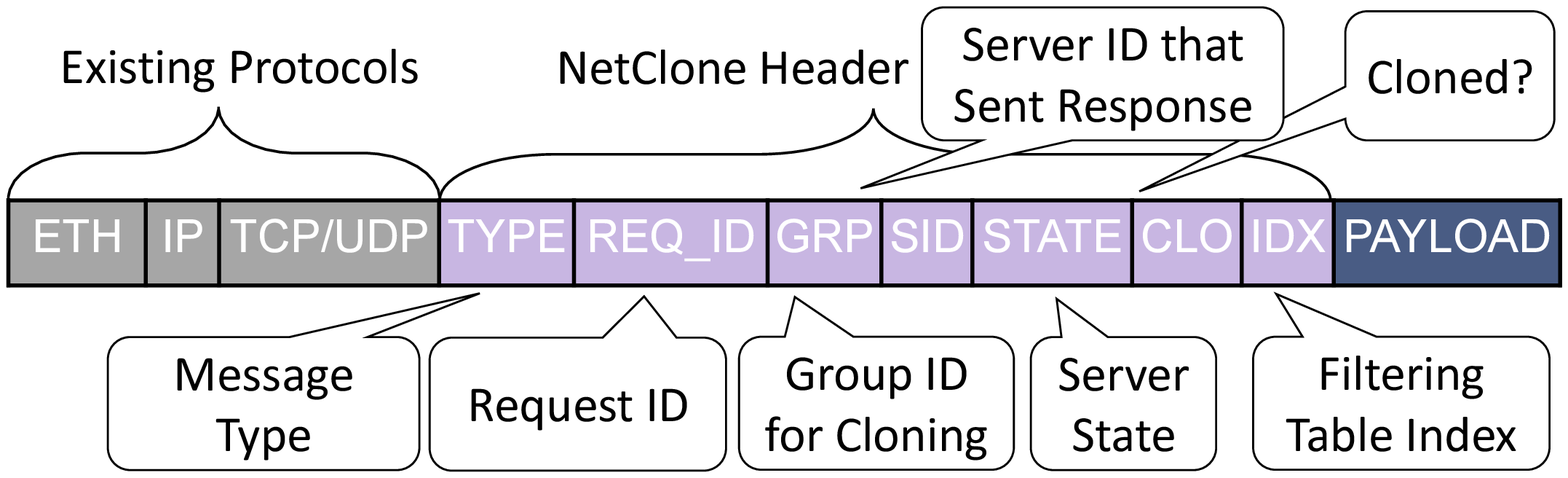}
\caption{\sys packet format. \label{fig:header}}
\end{figure}

\subsection{Packet Format}
Figure~\ref{fig:header} shows the packet format of \sys.
The \sys header is encapsulated as a L4 payload.
We reserve an L4 port number for \sys so that the switch can apply different packet processing logic for \sys packets and normal packets.
Since both \sys packets and normal packets are forwarded using traditional L3 routing, \sys is compatible with existing network functions.
The \sys header consists of 7 fields as follows.

\begin{itemize}[noitemsep]
\item{
\texttt{TYPE}: the message type, which can be \texttt{REQ} (a request) and \texttt{RESP} (a response).
}
\item{
\texttt{REQ\_ID}: the request ID, which is a unique sequence number assigned by the switch.
}
\item{
\texttt{GRP}: the group ID that specifies a pair of candidate servers.
}
\item{
\texttt{SID}: the server ID that sent a response.
This field is used as the index for the server state table.
}
\item{
\texttt{STATE}: the server state, which can be busy or idle.
}
\item{
\texttt{CLO}: the field to clarify whether the request is cloned or not. 0 means the non-cloned request; 1 means the cloned original request; 2 means cloned request.
}
\item{
\texttt{IDX}: the index for hash tables to filter redundant responses. 
Note that this is the table index, not the slot index of a table.
}
\end{itemize}

\begin{algorithm}[t!]
\small
\caption{Packet Processing in Data Plane \label{alg1}} 
\begin{algorithmic}[1] 
\StateX $-$ $pkt$: Packet to be processed
\StateX $-$ $SEQ$: Global sequence number for request IDs.
\StateX $-$ $GrpT$: Match-action table to get a server pair
\StateX $-$ $AddrT$: Match-action table to get IP address
\StateX $-$ $StateT$: Register array to track server states 
\StateX $-$ $ShadowT$: The copy of the state table
\StateX $-$ $FilterT$: Register arrays to filter redundant responses
\If{$pkt.type$ == \texttt{REQ} \textbf{and} NotCloned} 
    \State $SEQ \gets SEQ + 1$
    \State $pkt.req\_id \gets SEQ$ 
    \State $Srv1,Srv2 \gets GrpT.read(pkt.grp)$ \Comment{Get server IDs}
    \State $pkt.dst \gets AddrT[Srv1]$ \Comment{Get IP addr.}
    \If{$StateT[Srv1] == IDLE$ \textbf{and} $ShadowT[Srv2] == IDLE$} 

        \State $pkt.clo \gets 1$ \Comment{Mark as cloned original packet}
        \State $pkt.sid \gets Srv2$ \Comment{Will be used for forwarding clone}
        \State Clone($pkt$) \Comment{Forward $pkt$ and recirculate clone} 
    \EndIf
\ElsIf{$pkt.type$ == \texttt{REQ} \textbf{and} Cloned} 
        \State $pkt.clo \gets 2$ \Comment{Mark as cloned packet}
        \State $pkt.dst \gets AddrT[pkt.sid]$ \Comment{Get IP addr.}
\ElsIf{$pkt.type$ == \texttt{REP}} 
    \State $StateT[pkt.sid] \gets pkt.state$ 
    \State $ShadowT[pkt.sid] \gets pkt.state$ 
    \If{$pkt.clo > 0$} 
        \State $Hidx \gets Hash(pkt.req\_id)$ \Comment{Get hash index}
        \If{$FilterT[pkt.idx][HIdx] == pkt.req\_id$}
            \State $FilterT[pkt.idx][HIdx] \gets 0$ 
            \State Drop($pkt$) 
        \Else 
            \State $FilterT[pkt.idx][HIdx] \gets pkt.req\_id$ 
        \EndIf
    \EndIf

\EndIf
\State Forward($pkt$) 
\end{algorithmic}
\end{algorithm}

\subsection{Request Packet Processing}
\begin{figure}[t]
\centering
\subfloat[Request packets]{\includegraphics[width=\linewidth]{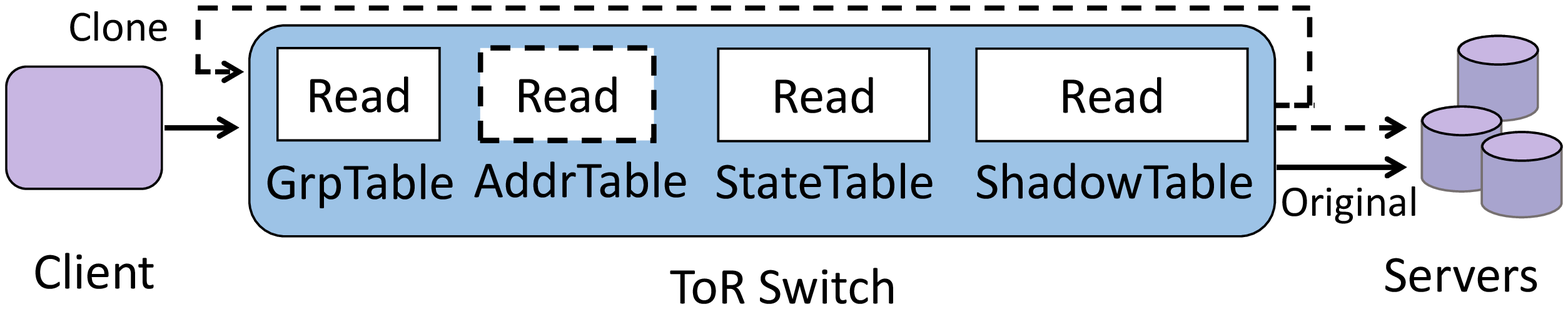}}\hfill
\subfloat[The faster response packet of a request]{\includegraphics[width=\linewidth]{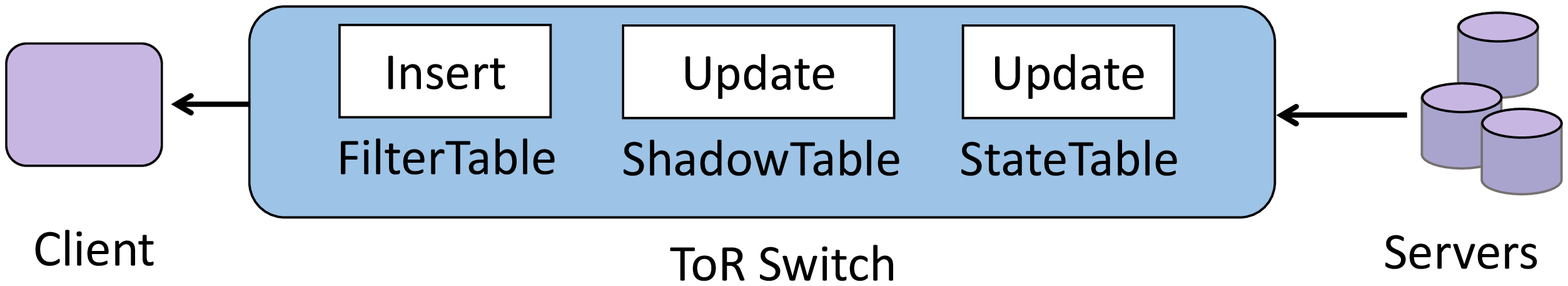}}\hfill
\subfloat[The slower response packet of the request]{\includegraphics[width=\linewidth]{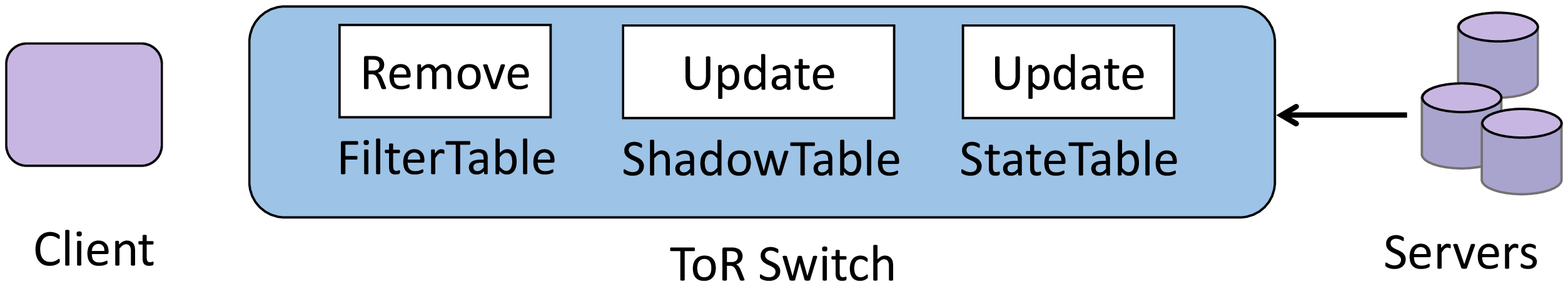}}
\caption{Request processing in \sys.
\label{fig:pro}}
\end{figure}

In this subsection, we describe how the switch processes request and response packets.
Algorithm~\ref{alg1} describes the high-level pseudocode of request processing in the switch data plane.
Figure~\ref{fig:pro} shows how \sys handles requests.

\textbf{Request packets.}
Clients do not have to know server information since the switch determines the destination server.
Clients use a group ID to determine a pair of candidate servers.
The group ID is randomly chosen by the client.
Each group ID matches two candidate servers, which are predefined by the operator.
The number of groups is $2\binom{n}{2}$ as we choose two servers between $n$ servers.
Multiplying by two is to sustain the randomness of server selection because the switch forwards the request to the first candidate server if cloning conditions are not satisfied.
For example, assume that we have only two servers.
In this case, we have two groups, and each group specifies \{Srv1,Srv2\} and \{Srv2,Srv1\}.
If we specify only one group (e.g., \{Srv1,Srv2\}) for this case, all non-cloned requests are forwarded to Srv1.

We use several tables to process requests as follows.
The group table $GrpT$ is a match-action table that maps from the group ID to the IDs of candidate servers.
Since clients do not specify the destination server initially, we also use the address table $AddrT$, a match-action table that assigns a destination IP address to the packet.
To track server states, we use two tables, which are the state table $StateT$ and the shadow state table $ShadowT$, a copy of $StateT$.
The tables contain the state of servers, and the switch performs cloning decisions based on the information.

The switch has different processing logic for original requests and cloned requests.
Upon receiving a normal request, the switch assigns a request ID to the request after increasing the sequence number by one (lines 1-3).
Next, the switch gets the ID of candidate servers (i.e., $Srv1$ and $Srv2$) by accessing $GrpT$ (line 4).
After that, the destination IP address is updated using the ID of server 1 as the index for $AddrT$ (line 5).
The switch now checks whether the tracked server states are both idle or not.
This is done by accessing $StateT$ and $ShadowT$ for server 1 and server 2, respectively (line 6).
If positive, the request is marked as cloned but original (line 7).
In addition, since we should forward the clone to server 2 as well, we put the ID of server 2 into the \texttt{SID} field (line 8).
The switch finally clones the request by forwarding the original request to server 1 and recirculating the cloned request into the ingress pipeline (line 9).
For recirculated cloned requests, the switch marks the request as cloned by updating the \texttt{CLO} field to 2 (lines 11-12).
After that, the IP address of server 2 is assigned to the request packet (line 13).
Figure~\ref{fig:pro} (a) outlines the process.

\textbf{Response packets.}
When a server sends back a response, the server updates the \texttt{SID} and \texttt{STATE} fields with its server ID and the current server state.
To handle responses, we use three tables: $StateT$, $ShadowT$, and $FilterT$.
Between them, $FilterT$ is the filter table to block slower responses, which is implemented as a register array.
The switch has slightly different logic for the faster response of a request and the slower response of the request, as shown in Figure~\ref{fig:pro} (b) and (c).
Upon receiving a response, the switch first updates the state information of the server in $StateT$ and $ShadowT$ (lines 14-16).
After that, the switch checks whether the response is of a cloned request by lookup the \texttt{CLO} field.
If positive, the switch data plane gets the hash slot index using the \texttt{REQ\_ID} field (lines 17-18).
If the hash slot of the matched filter table contains the same request ID (i.e., the slower response), the switch clears the slot and drops the packet (lines 19-21).
This is because the faster response is already forwarded to the client.
Otherwise, for the faster response, the switch puts the value of \texttt{REQ\_ID} field into the hash slot as a fingerprint to block the slower response (lines 22-23).

\begin{figure}[t!]
\centering
\includegraphics[width=8.0cm]{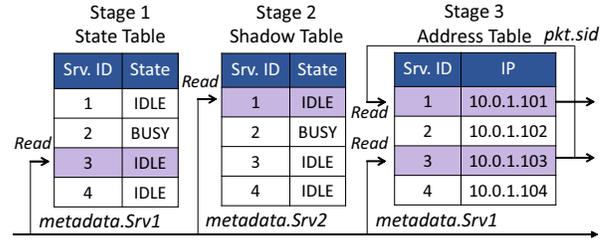}
\caption{Cloning decisions based on server states. \label{fig:cloning}}
\end{figure}

\begin{figure*}[t]
\centering
\subfloat[Faster response]{\includegraphics[width=0.330\linewidth]{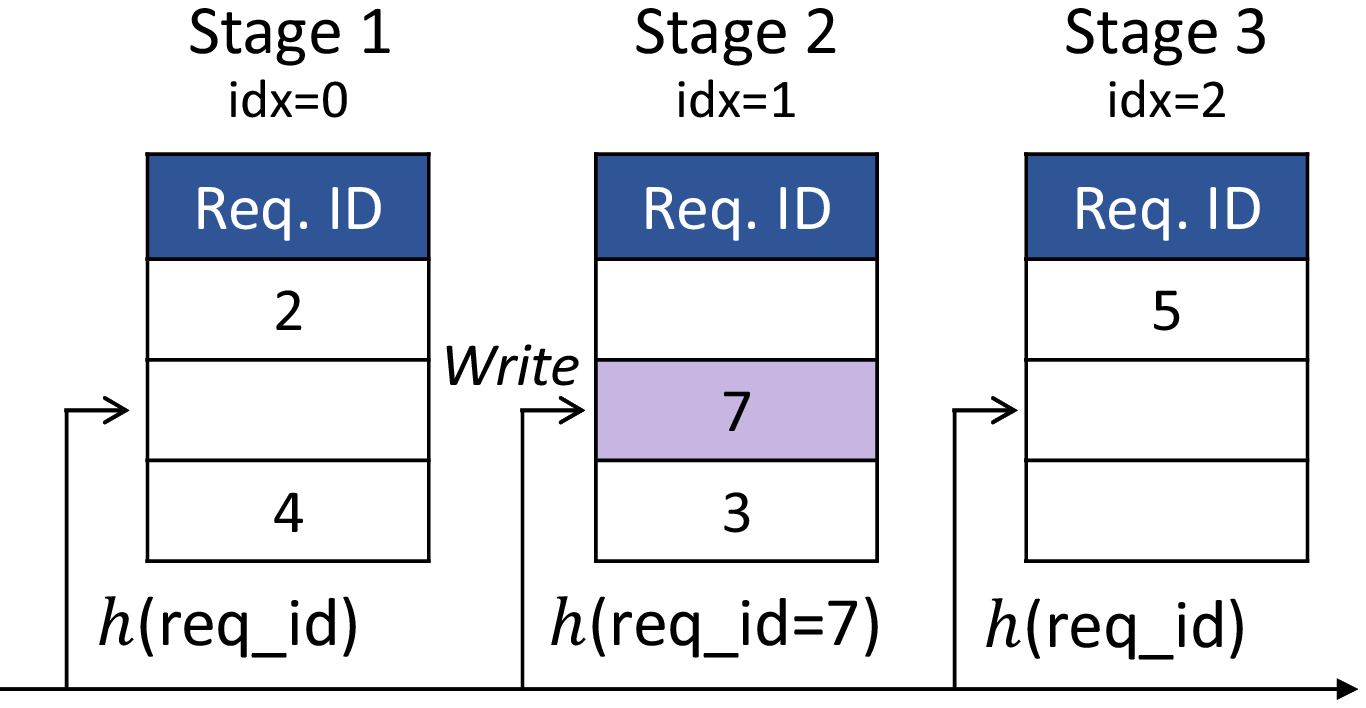}}\hfill
\subfloat[Slower response]{\includegraphics[width=0.330\linewidth]{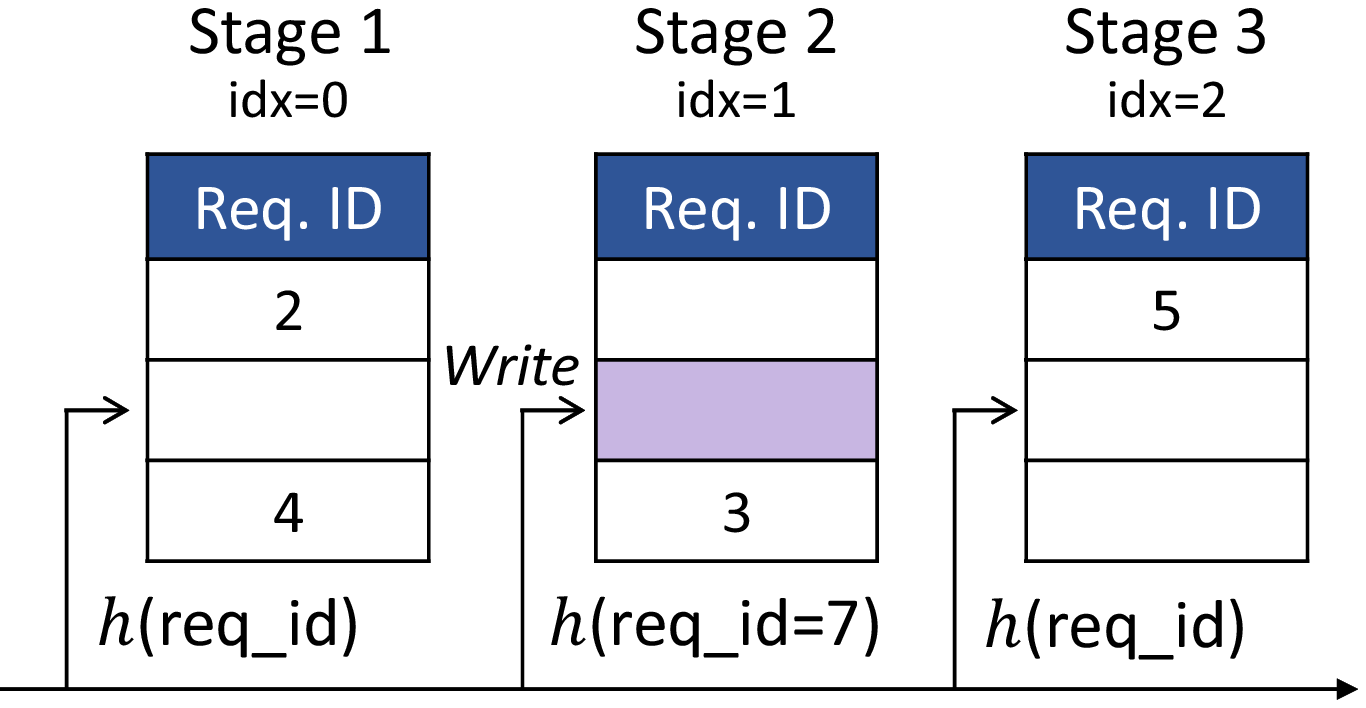}}\hfill
\subfloat[Handling hash collisions]{\includegraphics[width=0.330\linewidth]{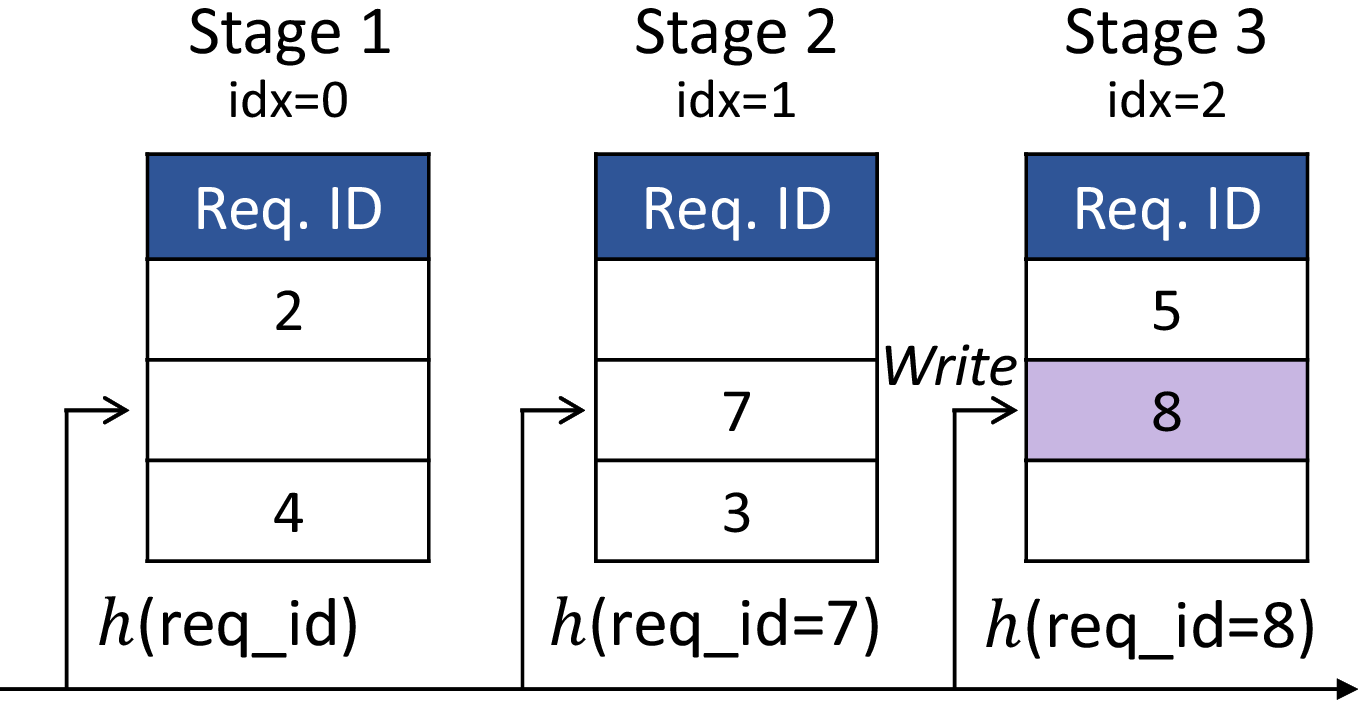}}\hfill
\caption{Examples of the filter table operations.
\label{fig:hash}}
\end{figure*}

\subsection{Dynamic Request Cloning}
\textbf{State tracking.}
\sota~\cite{laedge} dynamically replicates requests only if the candidate servers are idle, which is guaranteed by queueing requests in the coordinator and dispatching only one request at once.
However, this cannot be directly implemented in the current generation of programmable switches, as they have limited memory to buffer millions of requests and cannot store complex data structures.

Instead, we track server states and clone requests only if the servers are \textit{considered idle}.
We observe that, in general, the request queue in a server is empty if the number of incoming requests is not enough to make the server overloaded.
Therefore, if the queue is empty, we can consider the server as idle and is affordable to a cloned request as well.
We avoid non-empty queues because the existence of queued requests indicates that the server is too busy to handle incoming requests instantly, leading to performance degradation due to the reverse effect.
To deliver the server state to the switch, we make servers piggyback their state in response packets.
Upon receiving responses, the switch always updates the state table so that the latest state information can be maintained.

One issue is that the actual server state is not idle when a cloned request visits the server, because there is a time gap between the tracked state and the actual state.
Therefore, to avoid this, we make the server drop the packet request if the queue is not empty when receiving a cloned request.
It is important to note that only cloned requests (\texttt{CLO}=2) are dropped, while the original request (\texttt{CLO}=1) is processed normally.
Another possible solution is to track the server-side throughput and only clone requests when it is below a certain threshold.
However, this requires complex performance profiling to determine the threshold.

\textbf{Shadow state table}
To decide whether to clone, we must check the state of both candidate servers.
This requires the read of the state table twice, but it is not possible with the current switch ASIC using the PISA architecture~\cite{bosshart13}.
This is because, in the architecture, the switch data plane consists of multiple match-action stages and the memory space of the table is statically allocated to a single stage at compile time.
A packet passes through the match-action stages in the data plane to preserve a line-rate, which means that a packet can visit the table only once.
To overcome this limitation, we put a shadow table, a copy of the state table.
This allows the switch to check server states twice indirectly.
The consistency between the state table and the shadow table can be preserved since the switch always updates the tables at the same time upon receiving a response.

\textbf{Cloning in the switch.}
The commodity programmable switch provides two options to clone packets.
One is port mirroring and the other is multicasting.
Both of them generate a copy of the original packet and send the clone to a specific output port.
\sys utilizes multicasting since it is simpler in terms of switch configuration.
A challenge here is that we cannot assign the destination IP to the clone because the switch currently processes the original request at the time that replicates the request.
Therefore, we make the cloned request visit the ingress pipeline again using recirculation.
The recirculation is implemented by forwarding the clone to the port in loopback mode.
When the clone is recirculated, the switch assigns the destination IP address and forwards it to the corresponding output port.

\textbf{Example.} Figure~\ref{fig:cloning} shows an example of request cloning.
In this example, the switch confirms that the servers $Srv1$ and $Srv2$ are idle by accessing the state table and the shadow table.
The switch then assigns 10.0.1.103 to the request for the destination IP and forwards the request.
At the same time, the switch generates the cloned request for $Srv2$ and recirculates it.
When recirculated, the address table assigns 10.0.1.101 to the clone.
After that, the switch finishes the process by forwarding the clone.

\subsection{Response Filtering with Fingerprinting}
For the client, it is redundant to process the slower response of a request, since the client already received the faster response.
This overhead degrades the performance gain~\cite{vulimiri13}.
Our switch data plane filters redundant responses using request fingerprinting.
The idea is simple as follows.
We assign a monotonically-increasing sequence number for the request ID to each request, which is shared by the original and the clone.
The faster response puts the request ID in the filter table as a fingerprint to let the slower response know that the faster one is already processed.
The switch drops the slower response if the table contains the same request ID.
Note that the filter table is a register array, not a match-action table.

\textbf{Minimizing memory usage.}
A challenge here is how to minimize the memory footprint for response filtering because switch memory is a scarce resource.
However, the current switch ASIC does not allow dynamic memory allocation, and the memory space must be allocated at compile time~\cite{bosshart13,chole17}.
Reserving memory space as many as possible is not feasible because the sequence number for request IDs (i.e., the number of requests) can be over billions.

To reserve space for filtering while minimizing the memory footprint, we make the filter table use the hash index.
Our insight behind the idea is that the request ID only exists until the slower response arrives, which is a few microseconds in common.
Therefore, each hash slot can be reused for multiple request IDs.
We also allow responses to overwrite the existing request ID in a slot.
This is to handle hash collisions and packet drops.
If we prohibit the overwrite, responses can be dropped even if the response is the fastest one.
In a similar vein, if the slower response of a request is dropped or missed before visiting the switch, the hash slot becomes unavailable permanently.
Meanwhile, the overwrite may cause the failure to block the slower response, but it is not often since hash collisions and packet drops are rare because of microsecond-scale latency.

To further minimize hash collisions, we arrange multiple filter tables in the switch data plane.
We randomly assign a table index in the \texttt{IDX} field at the client side.
Since the \texttt{IDX} field remains consistent for a request and its responses, all related packets access the same table.
It is important to note that the index refers to the table index, not the hash slot index.
As a result, responses of two different requests with the same hash index can be processed concurrently without collisions, unless they have the same assigned table index.

\textbf{Example.}
Figure~\ref{fig:hash} shows examples of our filter table when we have three tables.
Let us consider a request with the request ID $req\_id=7$ and the table index $idx=1$.
As shown in Figure~\ref{fig:hash} (a), the faster response of the request inserts 7 to the empty hash slot in the second filter table.
When the slower response arrives at the switch, the switch resets the hash slot to empty and drops the response as illustrated in Figure~\ref{fig:hash} (b).
We now consider when a hash collision occurs in Figure~\ref{fig:hash} (c).
Although the hash index of the request with $req\_id=8$ collides with the request with $req\_id=7$, we can avoid the overwrite thanks to the different table index.

\subsection{Failure Handling}
In this section, we describe how \sys handles failures.

\textbf{Dropped messages.}
The loss of requests does not cause a problem as \sys simply concerns request cloning.
Meanwhile, the drop of responses may cause issues.
As mentioned in a previous subsection, if the slower response of a request is dropped, the filter table slot will be permanently occupied and unavailable.
However, our design allows responses to overwrite the hash slot with a different request ID, thus avoiding this problem.

\textbf{Server failures.}
In the event of a server failure, the overall performance will be degraded until the server is either recovered or removed.
The switch control plane can quickly remove the failed server from the list of potential destination servers by updating relevant tables (e.g., the group table and the address table) in the switch data plane and the number of groups on the client side.

\textbf{Switch failures.}
\sys does not cause any permanent misbehavior during switch failures as it stores only soft states, such as server states, the global sequence number for request IDs, and the filter table entries.
Once the switch is recovered, the server states can be updated through the following responses.
The loss of table entries does not lead to any serious consequence, although there may exist temporary overhead on the client side as slower responses can be forwarded.
Additionally, while the sequence number restarts from 0, this does not result in any fatal outcome, as most requests with earlier sequence numbers have already been completed.

\subsection{Handling Practical Requirements}
We now describe how \sys can support a variety of practical requirements.

\textbf{Integration with RackSched.}
RackSched~\cite{racksched} is an in-network request scheduler for microsecond-scale workloads.
It performs the Join-the-Shortest-Queue (JSQ) load balancing~\cite{randomloadbalancing} by utilizing the power of two choices~\cite{poweroftwochoices} in the switch data plane.
\sys can integrate RackSched into its design to make synergy as follows.
First, we change the state table to the load table and store the queue length of request queues in servers instead of binary states.
\sys still can make a cloning decision since we consider the server with the empty queue as idle.
If all candidate servers have empty queues, we replicate requests as usual.
Otherwise, we fall back to RackSched.
The switch compares the queue lengths of the servers and chooses the one with the shortest queue as the destination server.
When integrating the two solutions, we address several challenges caused by the computational limits of the switch ASIC, but we omit the detail due to space constraints.

\textbf{Multi-rack deployment.}
\sys generally targets a single-rack model like rack-scale computers, but it is possible to deploy for a multi-rack model like cloud-scale data centers.
This is because \sys leverages the existing forwarding function to route both normal and cloned requests.
In multi-rack deployment, aggregation switches do not have to be aware of request cloning and only ToR switches need to use \sys logic.
However, the ToR switch of servers may apply the \sys logic to packets even if the \sys processing should be done only in the ToR switch of the client.
Therefore, we add a switch ID field to the \sys header, with an initial value of zero that is set to the pre-defined switch ID when the packet passes through the ToR switch of the client.
ToR switches then apply the \sys logic only to packets with a switch ID field of zero or matching their own ID.

\textbf{Multiple pipelines and scalability.}
Modern programmable switches consist of multiple pipelines and each pipeline is connected to a number of ports.
For example, for a 64-port switch with 4 pipelines, 16 ports are assigned to each pipeline.
Each pipeline basically does not share its table entries, metadata, and registers.
Therefore, the solution has limited scalability with a limited number of ports (i.e., the number of servers) if it supports a single pipeline only.
\sys can work with multiple pipelines.
For example, \sys can work between the client with pipeline 0 and the server with pipeline 1.
This is because the \sys mechanism only requires soft states like server states and request IDs to be maintained in a pipeline connected to the client.
The entry of match-action tables of all pipelines can be updated by the switch control plane at the same time.
In a similar vein, \sys does not limit the number of supported servers compared to the vanilla switch even with multi-rack environments.
Our soft states are updated in the switch data plane at line-rate and do not rely on the switch control plane, which has a limited update throughput.

\textbf{Multi-packet messages.}
A microsecond-scale RPC message is generally small and consists of a single packet.
For example, DeathStarBench~\cite{deathstarbench} states that 75\% of RPC requests are less than 512 bytes in size, while over 90\% of RPC responses are smaller than 64 bytes.
Therefore, the current \sys design does not consider multi-packet requests and responses by default.
For multi-packet requests, since we assign the group ID at the client, the request affinity is naturally preserved.
However, we need a cloned request table that stores the ID of cloned but unfinished requests since every packet of a cloned request should be cloned regardless of system load.
To filter multi-packet responses, we can use multiple ordered filter tables and make the server assign a unique table index to each packet of a multi-packet response.
The switch then uses the corresponding ordered filter table index to perform the filtering logic.
For example, for a 4-packet response, the server could assign table indices from 0 to 3, and the switch would filter each packet using the matching filter table index.

\textbf{Protocol support.}
Our design basically considers UDP for the L4 protocol since our focus is on microsecond-scale RPCs, which usually consist of a single packet~\cite{alto,deathstarbench,racksched}.
However, some RPC applications may choose to use TCP.
To support TCP without any unexpected behavior, the request ID assignment logic should be revised.
This is because the switch will assign a different request ID for retransmitted requests, which can lead to misbehavior for multi-packet requests.
To address this, we use a tuple of the client ID and a local sequence number generated by the client for request IDs like Lamport clocks~\cite{lamportclock,hermes}.
Additionally, we also append the \sys header to the TCP handshake packet for applying the \sys logic though they do not contain a payload.

\textbf{Integration with RPC frameworks.}
We clarify that integrating \sys with existing RPC frameworks like eRPC~\cite{erpc} may require a considerable amount of engineering effort because the framework provides various functions and diverse packet transport logic like RDMA, which can cause a functional collision with \sys.
For example, some RPC frameworks can generate multiple packets for a single request even if the request size is small.
In this case, \sys may clone only partial packets of a multi-packet request when the server state information is updated.
This does not provide as much performance improvement as full cloning.
However, this still provides a degree of improvement since cloned packets of the request see the performance gain.
The retransmission of a request is a similar case since the tracked server state is continuously changed.
However, it is intentional that original packets and retransmitted packets may differ in copying or not because we should clone packets by considering the state of servers.
Furthermore, RPC frameworks should have a redundancy filtering mechanism or redundancy-aware response handling mechanism because a redundant response may not be filtered by the switch.
Without such a mechanism, RPC frameworks may not work correctly.
To support RDMA-based RPC frameworks, we need to revise the switch data plane partially for parsing the RDMA header and should address potential issues.
Note that it is possible to parse and craft RDMA packets in the switch data plane~\cite{kim20tea}.

\subsection{Generality}
A high-level takeaway from our work is that switches are an attractive high-performance vantage point to perform various functions in the era of microseconds.
Therefore, we believe that the design choice of \sys can inspire the emergence of other in-network computing systems for microsecond-scale applications.
Furthermore, the proposed techniques can be applied to various in-network computing systems.
For example, we leverage recirculation to assign the IP address to the cloned packet.
This can be applied to other systems that replicate packets in the programmable switch, which include data replication, consensus, and multi-path routing.
Request filtering using request fingerprints in the data plane can also be applied to other systems.
For example, an admission control mechanism that requires frequent and quick rule updates can utilize this.
This is because the rule update by the switch control plane is slow and has limited update throughput whereas the register update in the switch data plane is fast and offers line-rate throughput.

%% file: sections/implementation.tex
\begin{figure*}[t]
\centering
\subfloat[Exp(25)]{\includegraphics[width=0.250\linewidth]{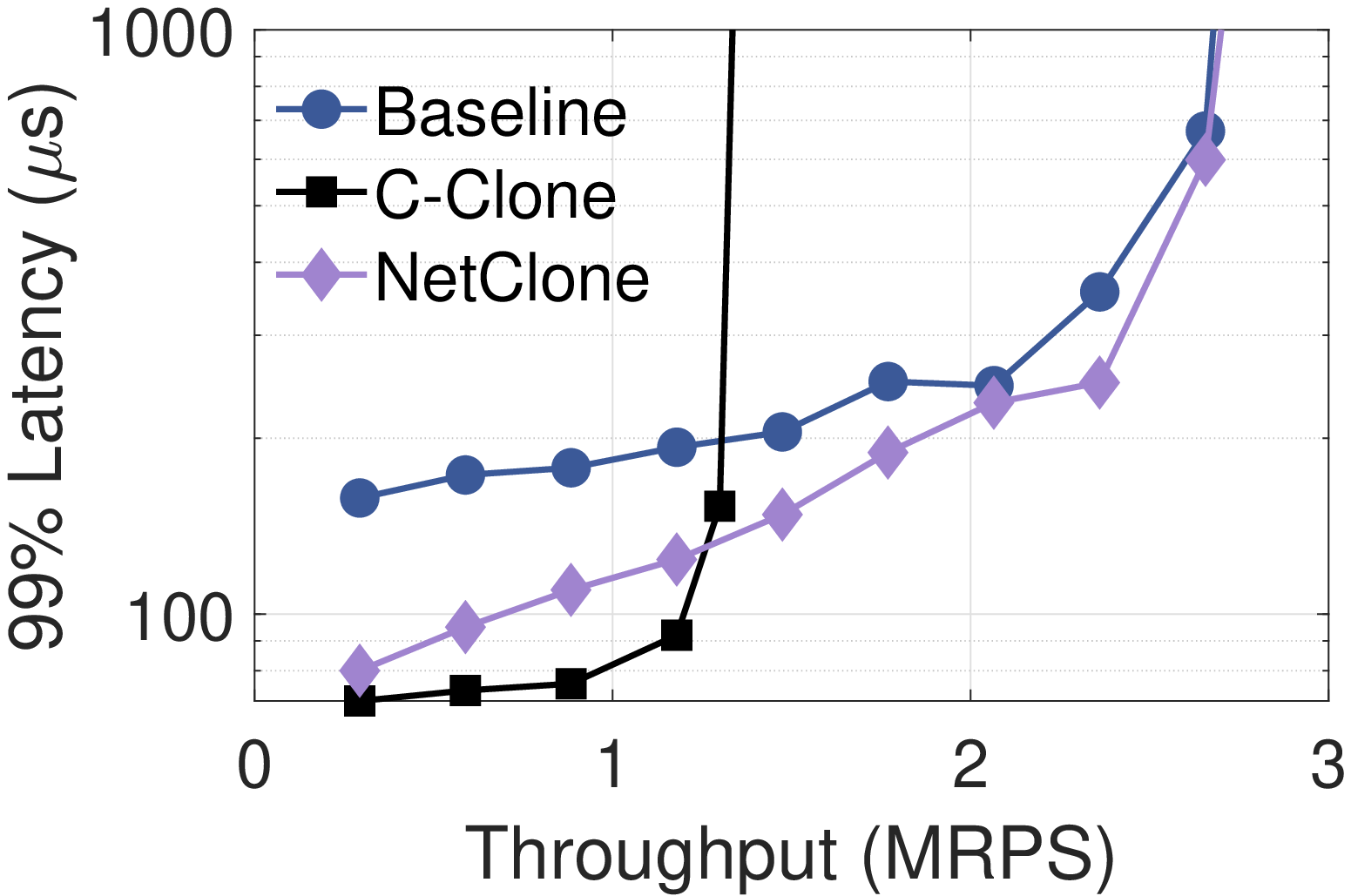}}\hfill
\subfloat[Bimodal(90\%-25,10\%-250)]{\includegraphics[width=0.250\linewidth]{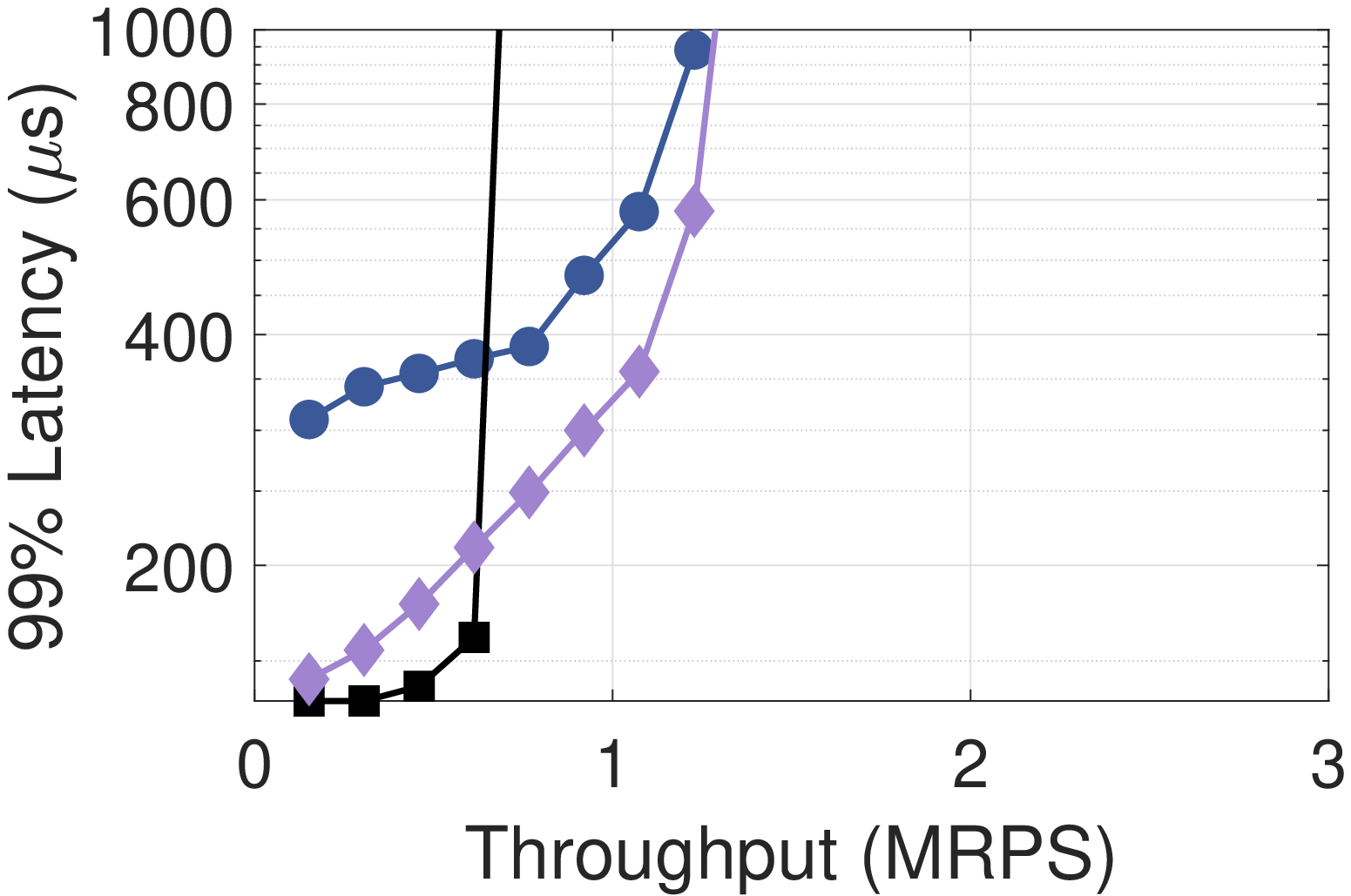}}\hfill
\subfloat[Exp(50)]{\includegraphics[width=0.250\linewidth]{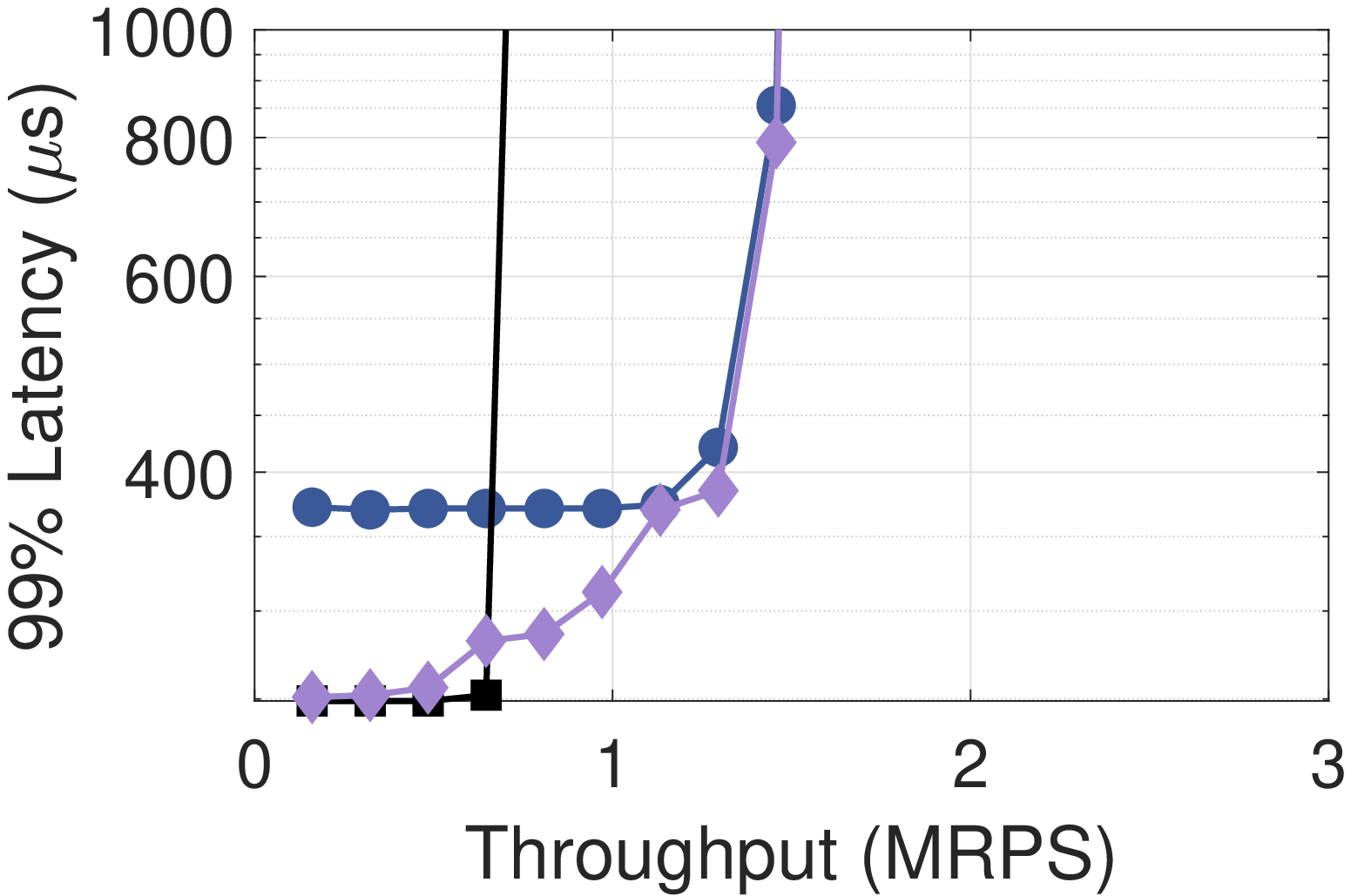}}\hfill
\subfloat[Bimodal(90\%-50,10\%-500)]{\includegraphics[width=0.250\linewidth]{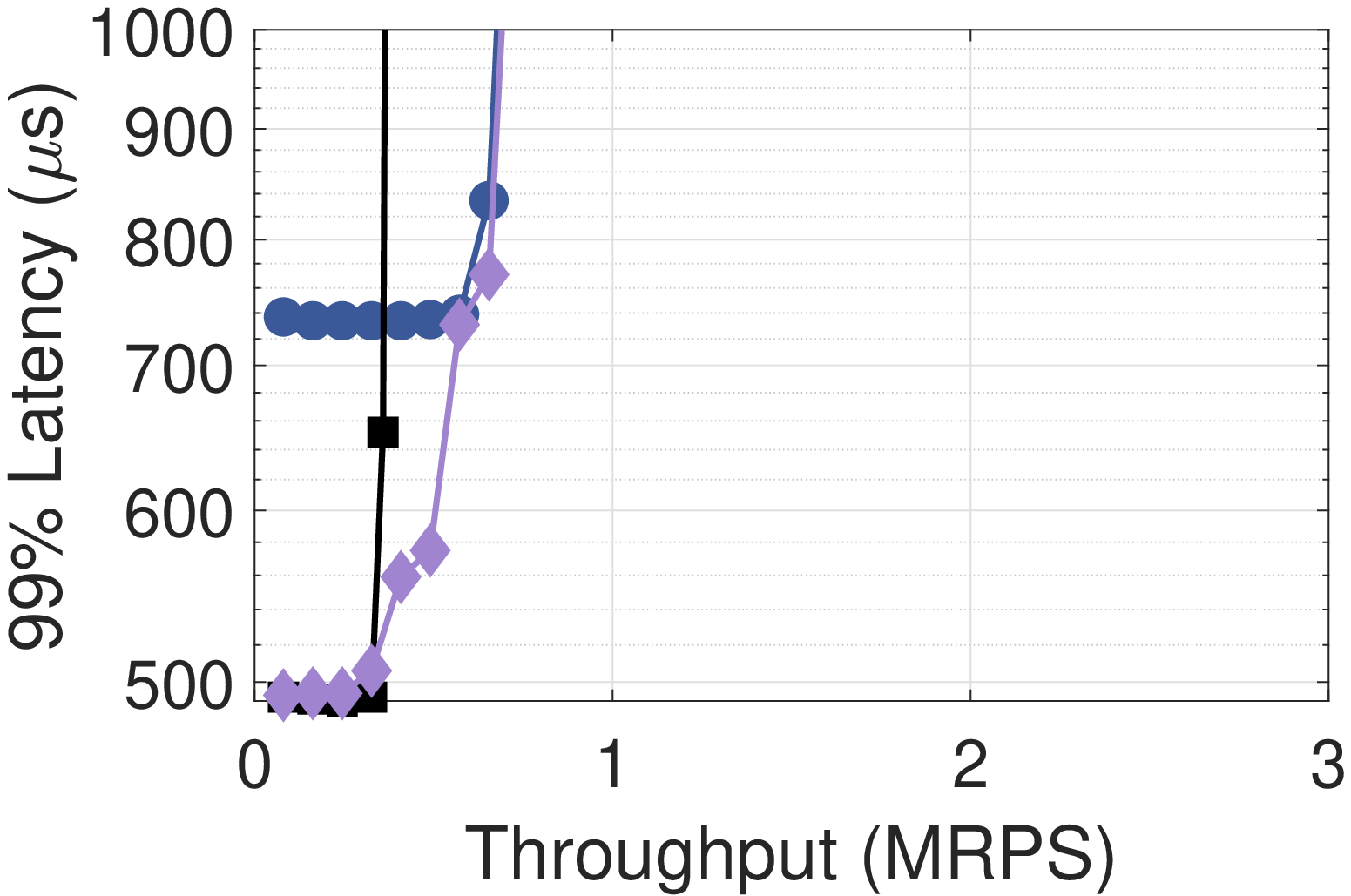}}\hfill
\caption{Experimental results for synthetic workloads.
\label{fig:main}}
\end{figure*}
\section{Implementation \label{implementation}}
\subsection{Switch Data Plane}
Our switch data plane is written in P4$_{16}$~\cite{bosshart14} and is compiled with Intel P4 Studio SDE 9.7.0 for Intel Tofino~\cite{tofinonodate}.
We implement our data plane modules in the ingress pipeline because the switch should finish cloning decisions before packet forwarding.
\sys consumes 7 match-action stages when using two filter tables.
We use 18.04\% SRAM, 12.28\% Match Input Crossbar, 26.79\% Hash Unit, and 21.43\% ALUs of the switch ASIC.
Most memory space is used to keep track of request IDs in the filter tables where each table has $2^{17}$ hash slots.
We can calculate how much throughput can be supported by the tables using a back-of-the-envelope calculation~\cite{racksched}.
When the average request latency is 50$\mu s$, each slot can handle 20 KRPS. 
As we have a total of $2^{18}$ slots, the current \sys prototype can support roughly 5.24 BRPS throughput.
Since we use a 32-bit slot, our hash tables use roughly 1.05 MB, which is 4.77\% of the switch memory.

\subsection{Client-Server Application}
We implement an open-loop multi-threaded application in C like prior work~\cite{racksched,breakwater,shinjuku}.
We use the NVIDIA Messaging Accelerator library (VMA)~\cite{vmanodate} for high-performance packet processing.
The VMA allows applications to process packets in userspace with RDMA-like kernel-bypass networking, minimizing the packet processing delay in hosts.
The client measures the throughput and latency by generating requests at a given target sending rate.
It consists of two threads, one is the sender thread and the other is the receiver thread.
The inter-arrival time between two consecutive requests is exponentially distributed.
The server consists of a single dispatcher thread and multiple worker threads.
The dispatcher enqueues received requests into a global request queue with FCFS policy.
Worker threads dequeue requests and process them in parallel.

%% file: sections/evaluation.tex
\section{Evaluation\label{evaluation}}

In this section, we evaluate \sys.
We first describe our experiment methodology.
Next, we present experimental results with various workloads and system conditions.

\subsection{Methodology}
\subsubsection{Testbed setup}
To evaluate \sys, we use a cluster consisting of 8 commodity servers, which are connected by an APS Networks BF6064X-T switch.
The switch data plane is based on a 6.5 Tbps Intel Tofino switch ASIC~\cite{tofinonodate}.
The servers are equipped with a 10-core CPU (Intel i5-12600K @ 3.7 Ghz, 12 hyperthreads and 4 non-hyperthreads), 32 GB of DDR5 memory, and a single-port 100GbE RDMA-capable NIC.
The servers run Ubuntu 20.04 LTS with Linux kernel 5.15.0.
Unless specified, 2 servers act as clients to generate requests and the remaining 6 servers are used as worker servers.
The performance bottleneck is at worker servers.

\subsubsection{Workloads}
We use a variety of synthetic and real-world application workloads similar to recent works~\cite{racksched,shinjuku,zygos}.
The workloads use one-packet requests and responses with UDP like RackSched~\cite{racksched}.

With a synthetic workload, a worker server processes a dummy RPC for a duration that we specify. 
The synthetic workload allows us to evaluate the performance of \sys with various applications by emulating any target distribution of services and variability.
Unless specified, we consider an exponential distribution with mean = 25 $\mu s$ by default, which can represent common short-lasting RPCs.
We also consider a bimodal distribution where 90\% are 25 $\mu s$ and 10\% are 250 $\mu s$, which represents a mix of simple and complex RPCs.
To inspect the impact of RPC duration, we use 50 $\mu s$ and 500 $\mu s$ as well.
To emulate the service-time variability, we follow the observations from \sota~\cite{laedge}.
We consider $p=0.01$ and $p=0.001$ to represent a high variability and a low variability, where $p$ denotes the jitter probability to experience excessive long latency.
We basically consider that workloads have high service-time variability, and the runtime of an RPC experiencing the unexpected jitter can take 15 times more than the normal case.
For the real-world application workload, we use Redis~\cite{redis}, a widely deployed in-memory key-value store in many production systems.

\subsubsection{Compared Schemes}
We compare our work against the baseline, \cc, \sota~\cite{laedge}.
The baseline sends requests to workers randomly without cloning.
\cc is the client-based cloning mechanism that always sends duplicate requests to two random worker servers.
\sota performs dynamic cloning using the coordinator.
In most experiments, we compare \sys to the baseline and \cc because \sota has significantly lower throughput than \sys.

\begin{figure}[t!]
\centering\hfill
\subfloat[Exp(25)]{\includegraphics[width=0.48\linewidth]{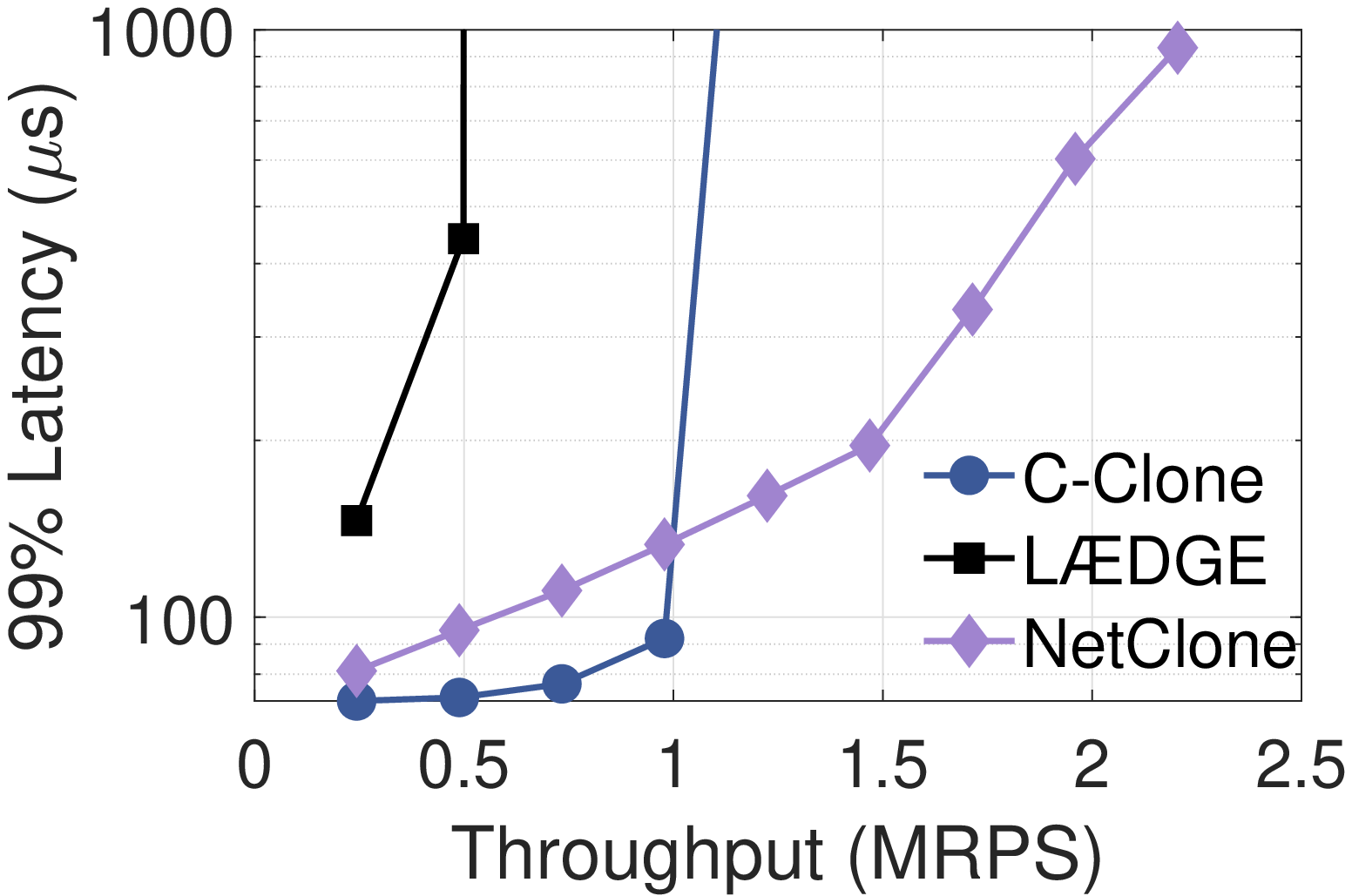}}\hfill
\subfloat[Bimodal(90\%-25,10\%-250)]{\includegraphics[width=0.48\linewidth]{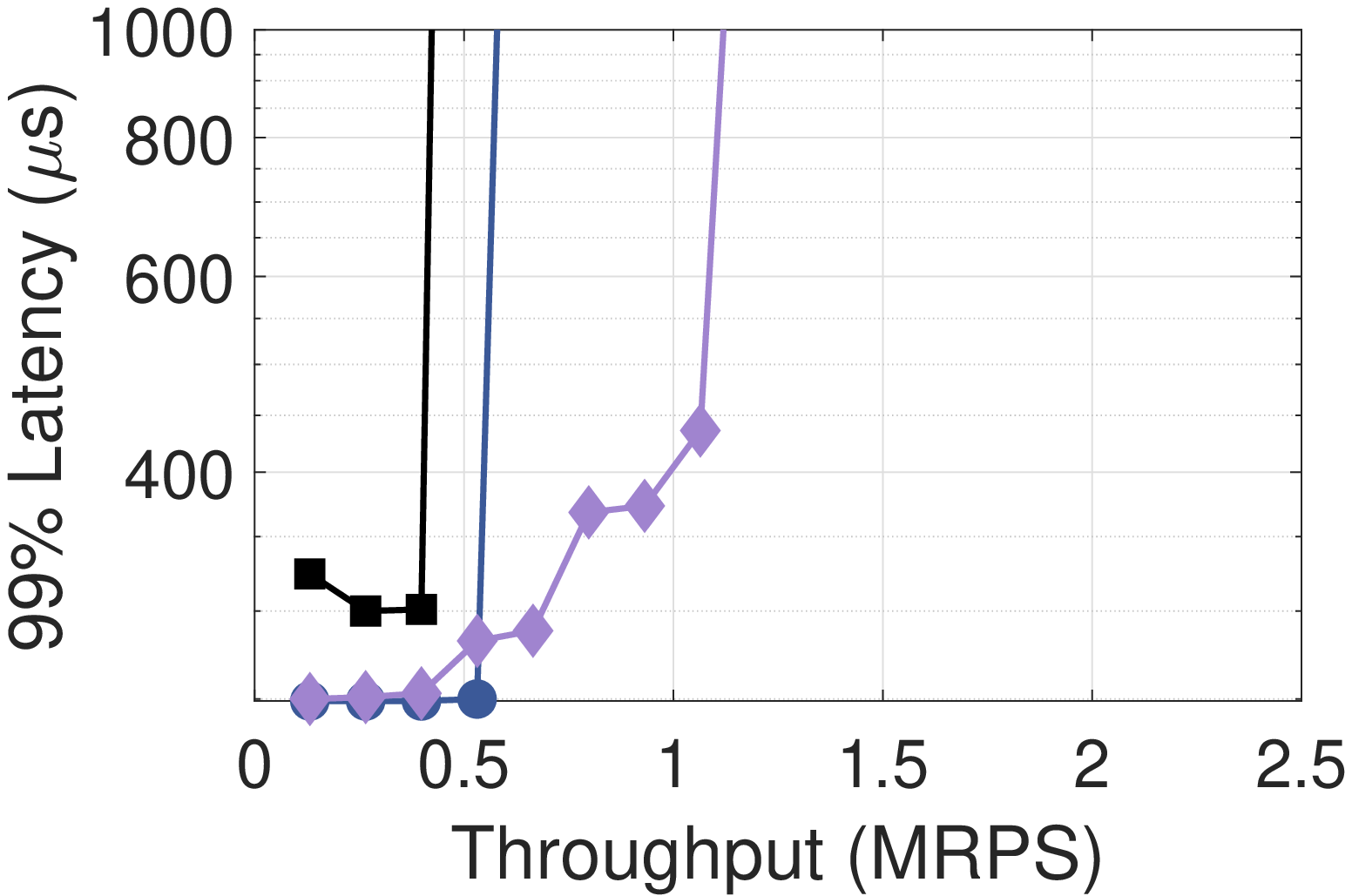}}\hfill
\caption{Comparison with the existing solutions.
\label{fig:laedge}}
\end{figure}

\subsection{Synthetic Workloads Results}
We plot the performance of three schemes, the baseline, \cc, and \sys in Figure~\ref{fig:main} for different workloads.
Note that Y-axis is in log scale for better visibility.
\cc shows limited throughput in all figures due to its static request cloning, which overloads worker servers beyond a certain point.
Thanks to the dynamic cloning and response filtering, \sys achieves low tail latency while maintaining similar throughput to the baseline.
In Figure~\ref{fig:main} (a) and (b), we can find that \sys achieves better latency than the baseline across almost all loads.
The average improvement is 1.48$\times$ and 1.27$\times$ for Exp(25) and Bimodal(90\%-25,10\%-250), respectively.
Since the work servers become busier as throughput grows, \sys clones requests less as well. 
Therefore, the degree of improvement decreases as the system load grows.
Meanwhile, at low loads, \sys experiences worse latency than \cc.
This occurs because \sys does not replicate requests when the tracked queue length is not zero.
Note that the queue can build up occasionally, even at low loads.
In Figure~\ref{fig:main} (c) and (d), \sys provides low tail latency at low loads, similar to Figure~\ref{fig:main} (a) and (b).
However, the performance improvement at high loads is negligible due to the longer processing time of RPCs that keeps the queue length non-zero at high loads.

\begin{figure}[t!]
\centering
\includegraphics[width=8.0cm]{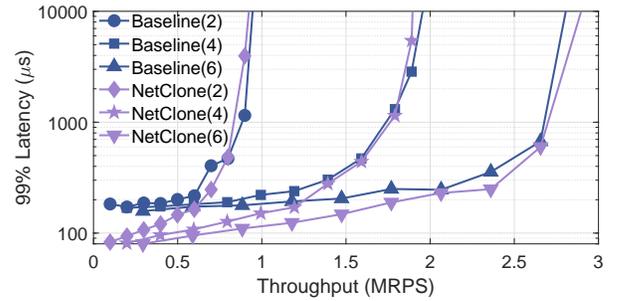}
\caption{Impact of the number of servers. \label{fig:scalability}}
\end{figure}

\subsection{Scalability}
\subsubsection{Comparison with the existing solutions}
In this experiment, we compare \sys with \cc and \sota to show that \sys has better throughput and scalability.
We use five worker servers because one server should be dedicated to the \sota coordinator.
The results, shown in Figure~\ref{fig:laedge}, indicate that \sys provides high throughput, while \sota and \cc exhibit low throughput.
\cc does not incur latency overhead to clone requests but its static cloning limits system throughput.
\sota performs even worse than \cc since it relies on a CPU-based coordinator to clone requests.
The coordinator server easily becomes a performance bottleneck, making it difficult to support high request rates with multiple worker servers.
Even with a highly optimized coordinator, \sota would be still behind \sys, as switches can process billions of packets per second, while optimized servers can handle only a few million packets per second.
This result demonstrates that performing request cloning in the switch is a desirable approach to achieve high performance.

\begin{figure*}[t!]
\centering\hfill
\subfloat[Exp-Homogeneous]{\includegraphics[width=0.250\linewidth]{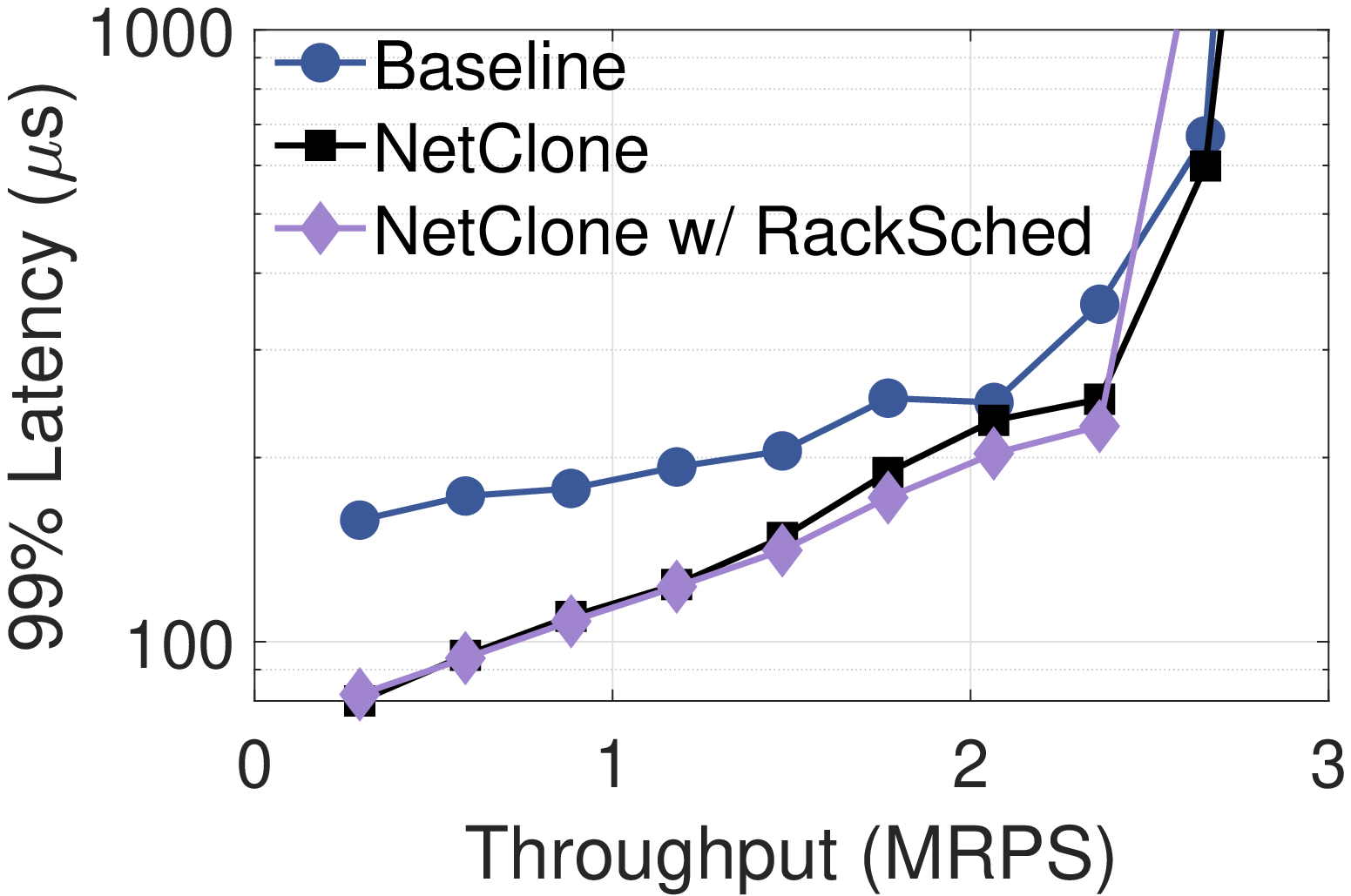}}\hfill
\subfloat[Exp-Heterogeneous]{\includegraphics[width=0.250\linewidth]{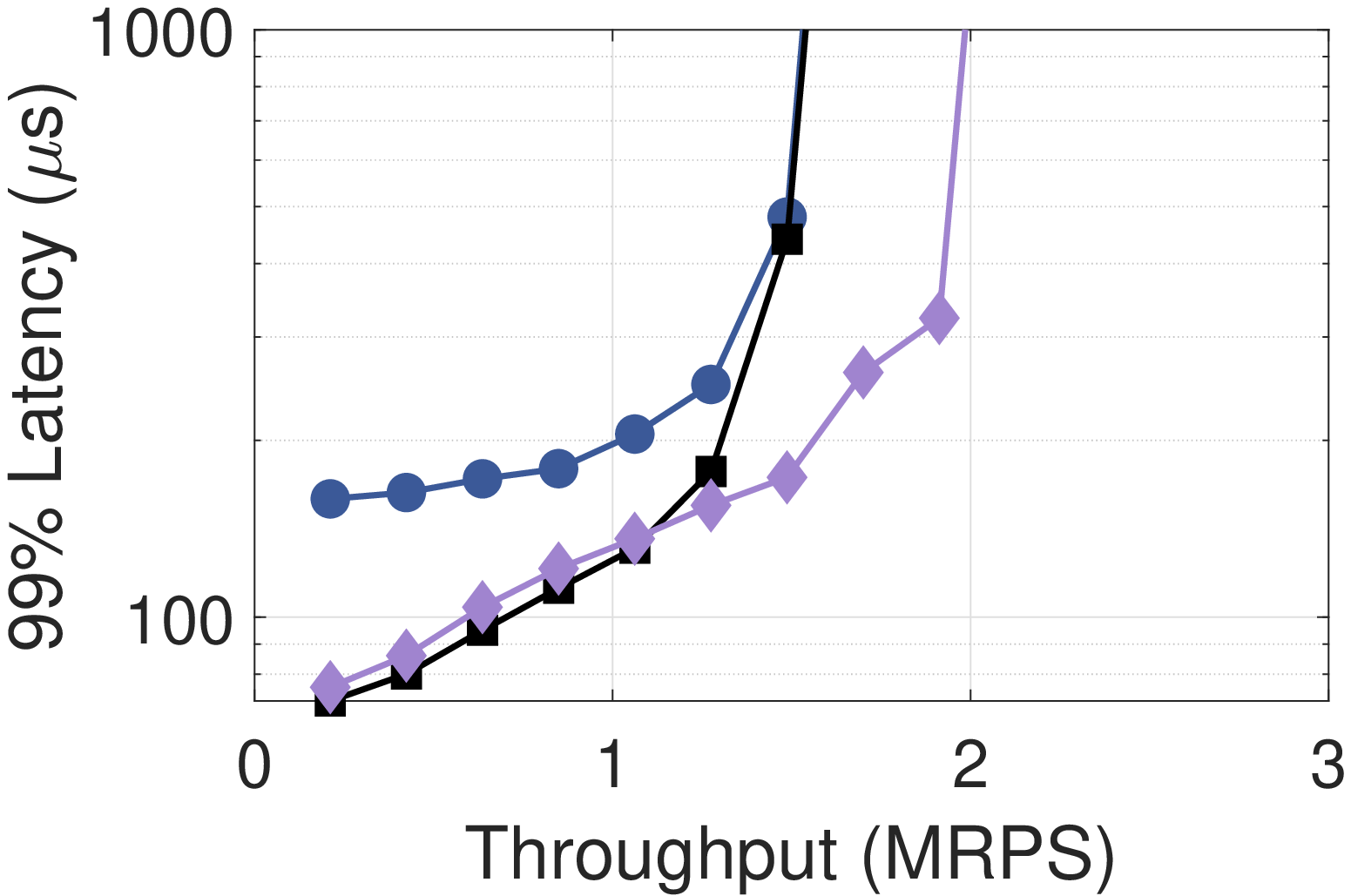}}\hfill
\subfloat[Bimodal-Homogeneous]
{\includegraphics[width=0.250\linewidth]{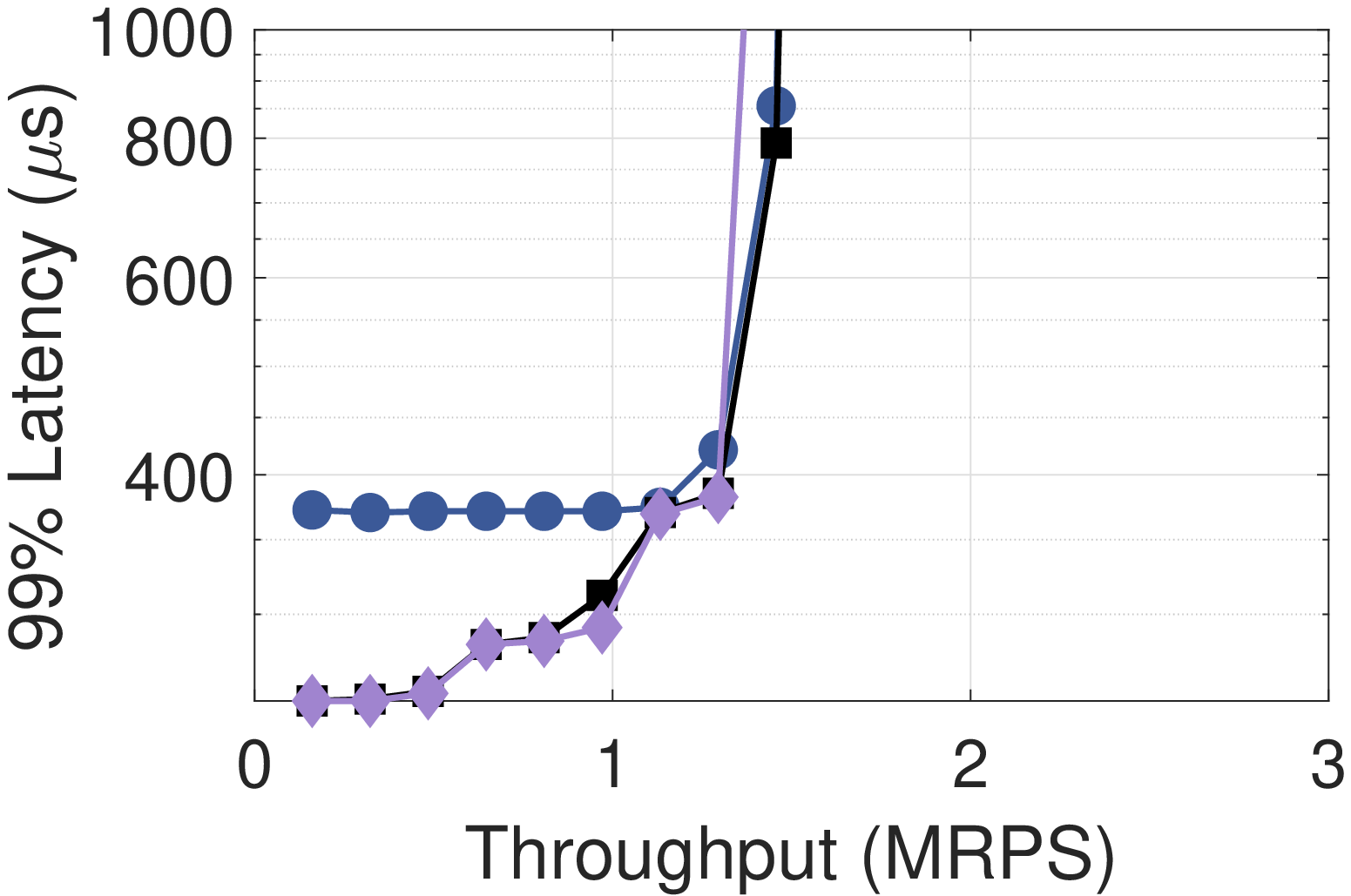}}\hfill
\subfloat[Bimodal-Heterogeneous]{\includegraphics[width=0.250\linewidth]{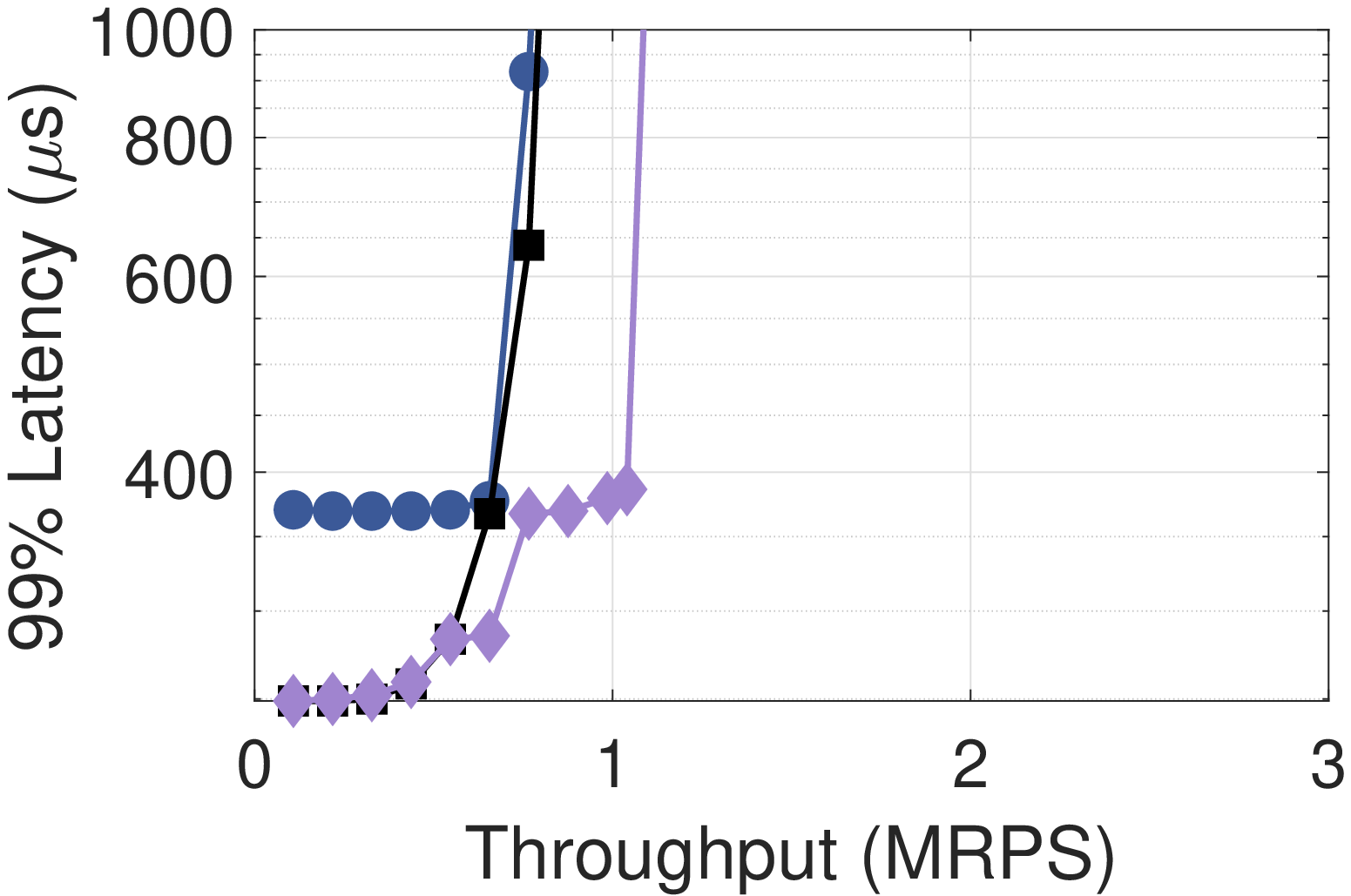}}\hfill
\caption{Performance with RackSched under homogeneous and heterogeneous workloads.
\label{fig:homo}}
\end{figure*}

\subsubsection{Impact of the number of servers}
We now evaluate the scalability of \sys by varying the number of worker servers.
As \sys performs request cloning in switches, it can scale out to multiple servers while maintaining low tail latency. Figure~\ref{fig:scalability} shows the results for 2, 4, and 6 worker servers.
We did not conduct an experiment with one server as \sys requires a minimum of two servers for redundancy.
As the number of worker servers increases, both \sys and the baseline show improved throughput.
\sys maintains lower tail latency than the baseline regardless of the number of servers.
One observation worth mentioning is that when the number of worker servers is two or four, \sys shows worse latency at very high loads.
This can be attributed to two reasons.
First, \sys sends cloned requests only when the server is idle, but the server may be busy in fact.
We drop cloned requests if the actual state is busy, but the processing cost can be harmful if the number of redundant requests is large at very high loads.
Second, with a small number of servers, there may not be enough idle servers available.
Therefore, many cloned requests are forwarded to actually overloaded servers for a short time with herding effects, resulting in high tail latency at very high loads.
However, when the number of servers is large, the probability of performance degradation decreases as \sys has a larger pool of servers to choose.

\begin{figure}[t!]
\centering\hfill
\subfloat[99\%-GET,-1\%-SCAN]{\includegraphics[width=0.500\linewidth]{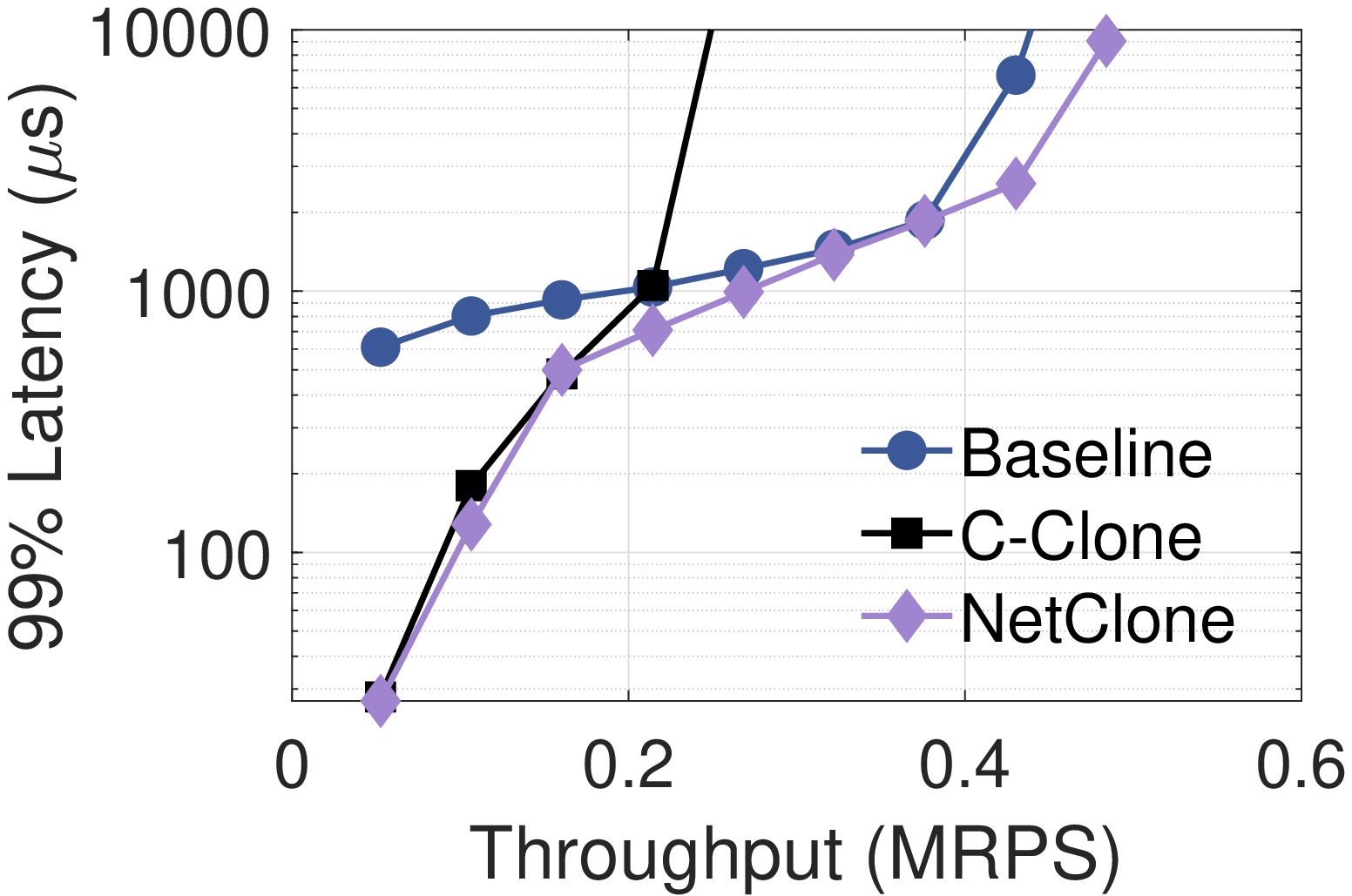}}\hfill
\subfloat[90\%-GET,10\%-SCAN]{\includegraphics[width=0.500\linewidth]{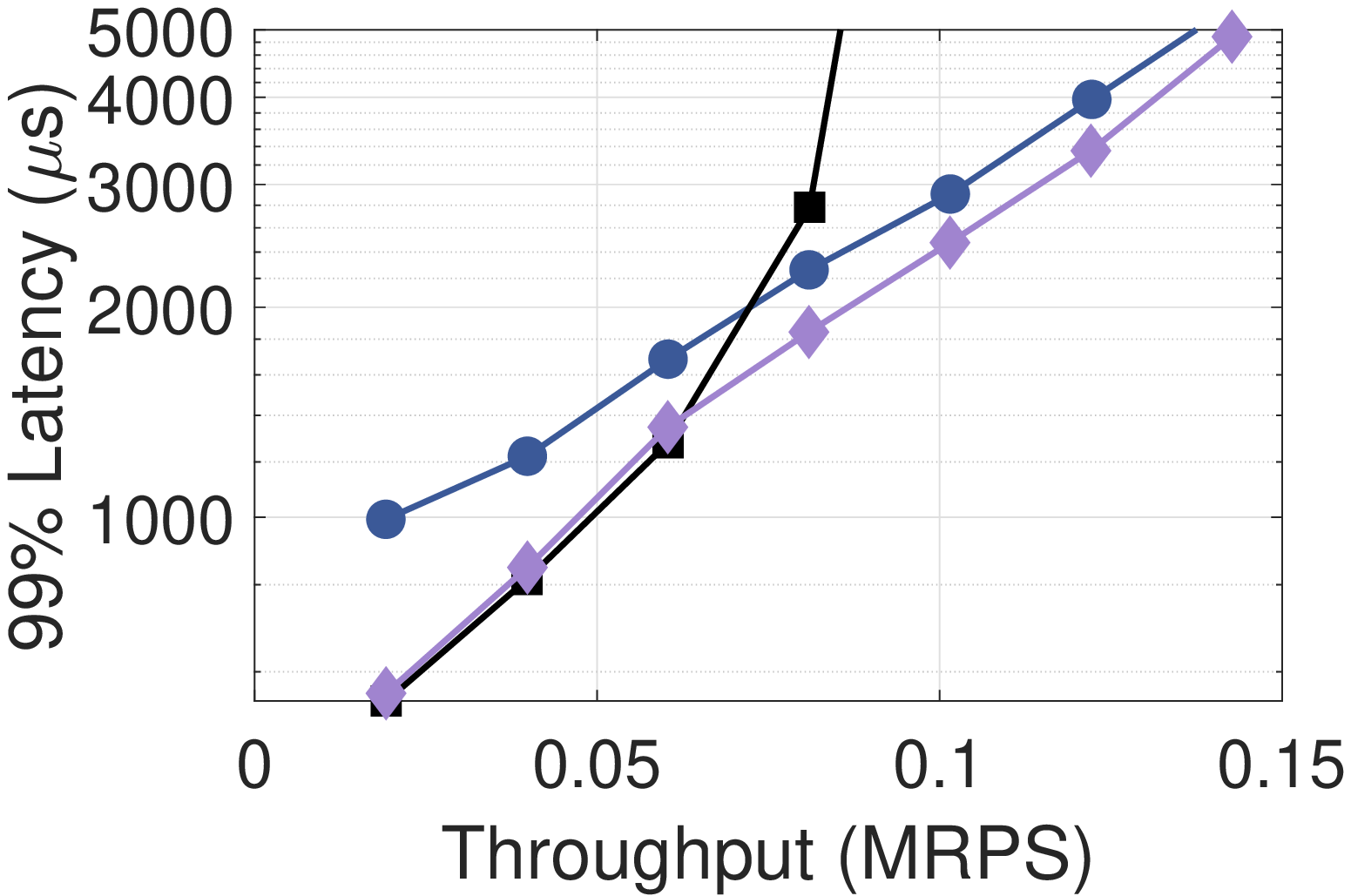}}\hfill
\caption{Experimental results for Redis.
\label{fig:redis}}
\end{figure}

\subsection{Performance with RackSched}

We now show how \sys can make synergy with Racksched~\cite{racksched}.
\sys contributes to reducing latency and RackSched is effective to improve throughput.
Figure~\ref{fig:homo} is experimental results for Exp(25) and Bimodal(90\%-25,10\%-250) workloads with a different number of workers.
The homogeneous workloads assume that each worker server has an equal number of worker threads (15 worker threads and 1 dispatcher thread).
In the heterogeneous workloads, three of the worker servers have 15 worker threads, while the other three have 8 worker threads.
We see that \sys with RackSched achieves the best performance, thanks to RackSched's ability to handle possible load imbalances between worker servers.
\sys with RackSched performs better with heterogeneous workloads than with homogeneous workloads because the latter workloads result in more imbalance loads.
Meanwhile, in homogeneous workloads, \sys with RackSched is worse than \sys at very high loads, and we suspect that this is because the cases when the tracked state and the actual state are unmatched increase as RackSched makes the request queue empty more often.

\begin{figure}[t!]
\centering\hfill
\subfloat[99\%-GET,-1\%-SCAN]{\includegraphics[width=0.500\linewidth]{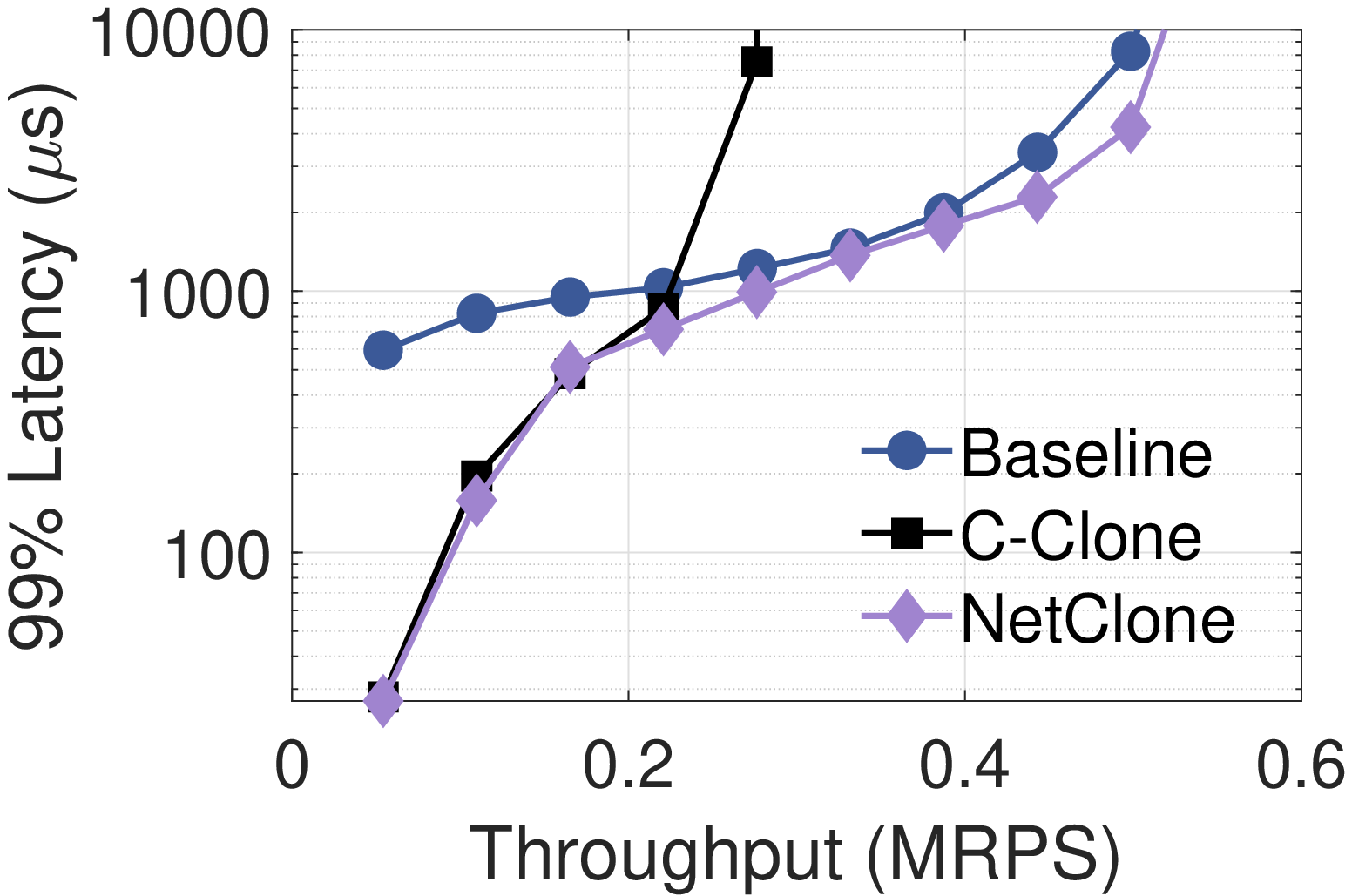}}\hfill
\subfloat[90\%-GET,10\%-SCAN]{\includegraphics[width=0.500\linewidth]{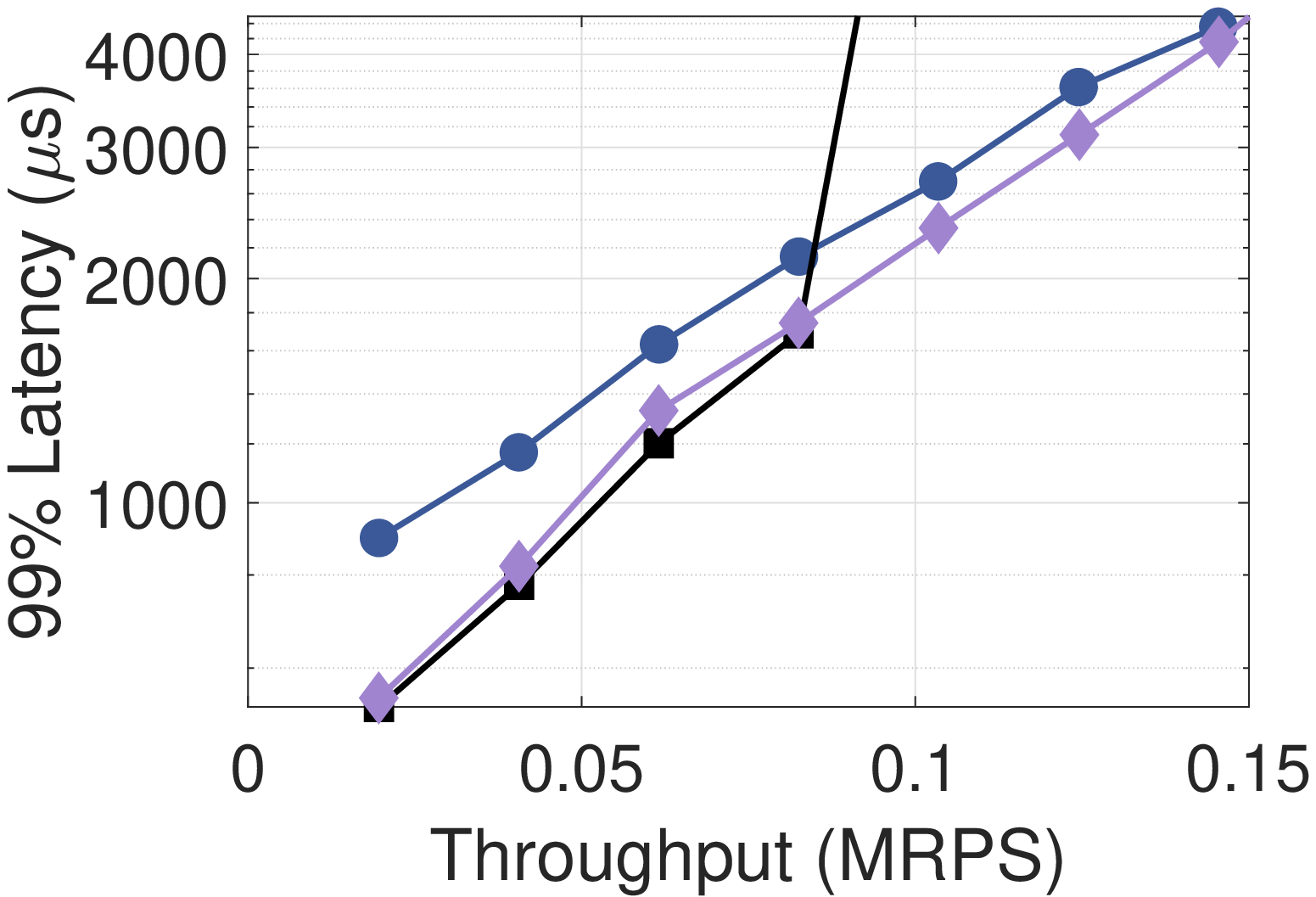}}\hfill
\caption{Experimental results for Memcached.
\label{fig:memcached}}
\end{figure}

\subsection{Applications: Redis and Memcached}
We now show that \sys is effective with real-world applications using Redis~\cite{redis} and Memcached~\cite{memcached}, which are popular in-memory key-value stores, commonly used in production services.
We conduct experiments using 1 million objects with 16-byte keys and 64-byte values~\cite{mica} by considering replicated key-value storage.
Unlike previous in-network solutions for key-value stores~\cite{jin17,netlr,jin18,harmonia}, \sys does not impose any limitations on the key or value sizes, as it does not store keys or values in the switch data plane.
In this experiment, clients generate read requests, and worker servers return values with a skewed key access pattern with Zipf-0.99.
Note that \sys does not clone write requests because the write coordination should be handled by replication protocols.
We use 8 worker threads in each worker server.
We vary the portion of GET and SCAN requests to 99\%-GET,1\%-SCAN and 90\%-GET,10\%-SCAN where GET reads a single object and SCAN reads 100 objects.

Figure~\ref{fig:redis} and Figure~\ref{fig:memcached} show results, which have similar trends.
We can see that, like the result with the synthetic workload, \sys improves tail latency by masking service-time variability.
The performance gap is the biggest at low loads, and the gap becomes small as throughput grows.
\cc shows similar tail latency to \sys, but its throughput is limited to half of \sys as expected.
In the Redis experiment, \sys is better than the baseline by up to 22.59$\times$ and 1.77$\times$ for 99\%-GET,1\%-SCAN and 90\%-GET,10\%-SCAN, respectively.
In Memcached, the largest improvement degree is 22.00$\times$ and the smallest one is 1.06$\times$ for 99\%-GET,1\%-SCAN.
For 90\%-GET,10\%-SCAN in Memcached, \sys achieves better tail latency than the baseline by 1.24$\times$ on average.

\begin{figure}[t]
\centering
\subfloat[Portion of empty queues]{\includegraphics[width=0.500\linewidth]{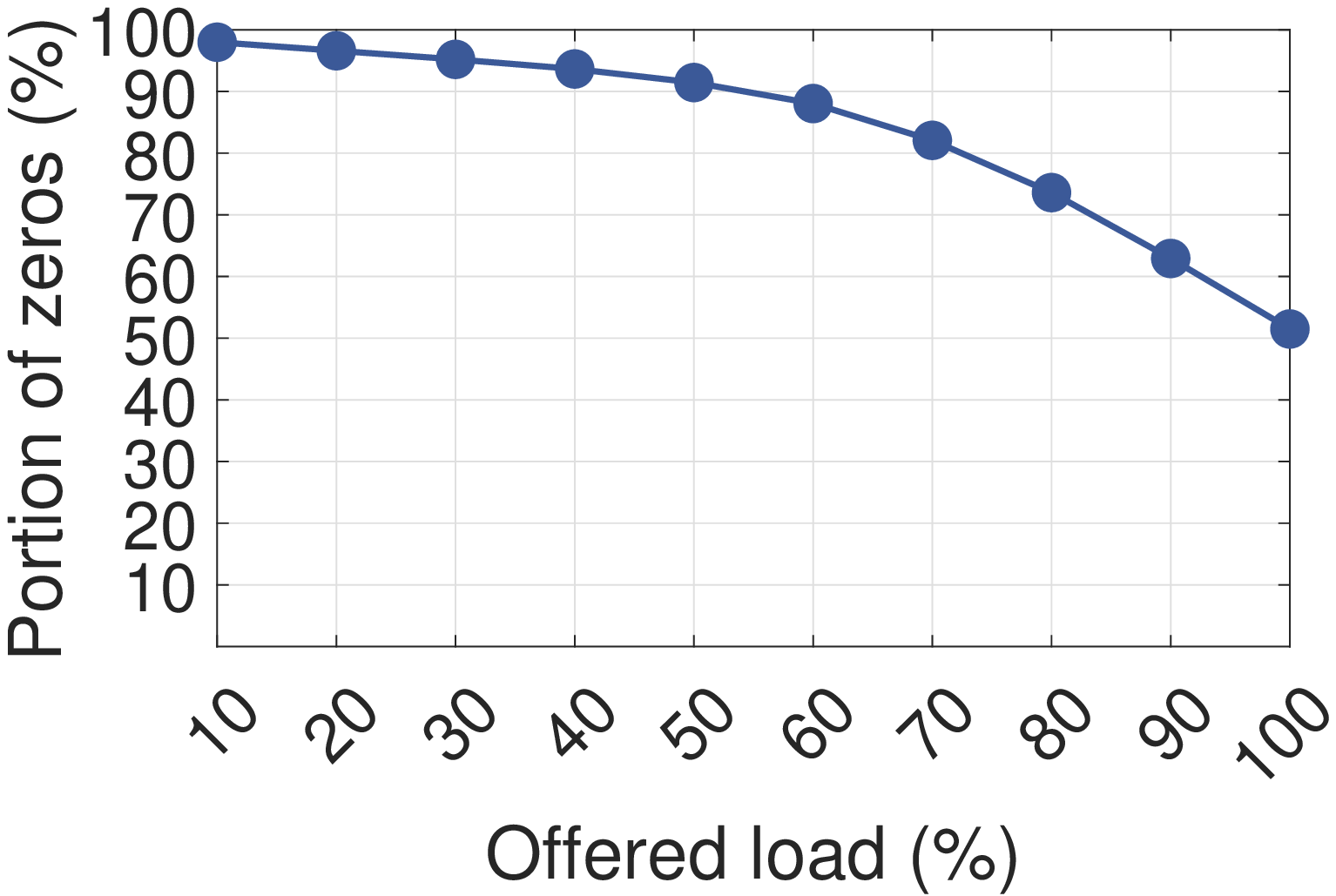}}\hfill
\subfloat[Latency at 90\% of load]{\includegraphics[width=0.500\linewidth]{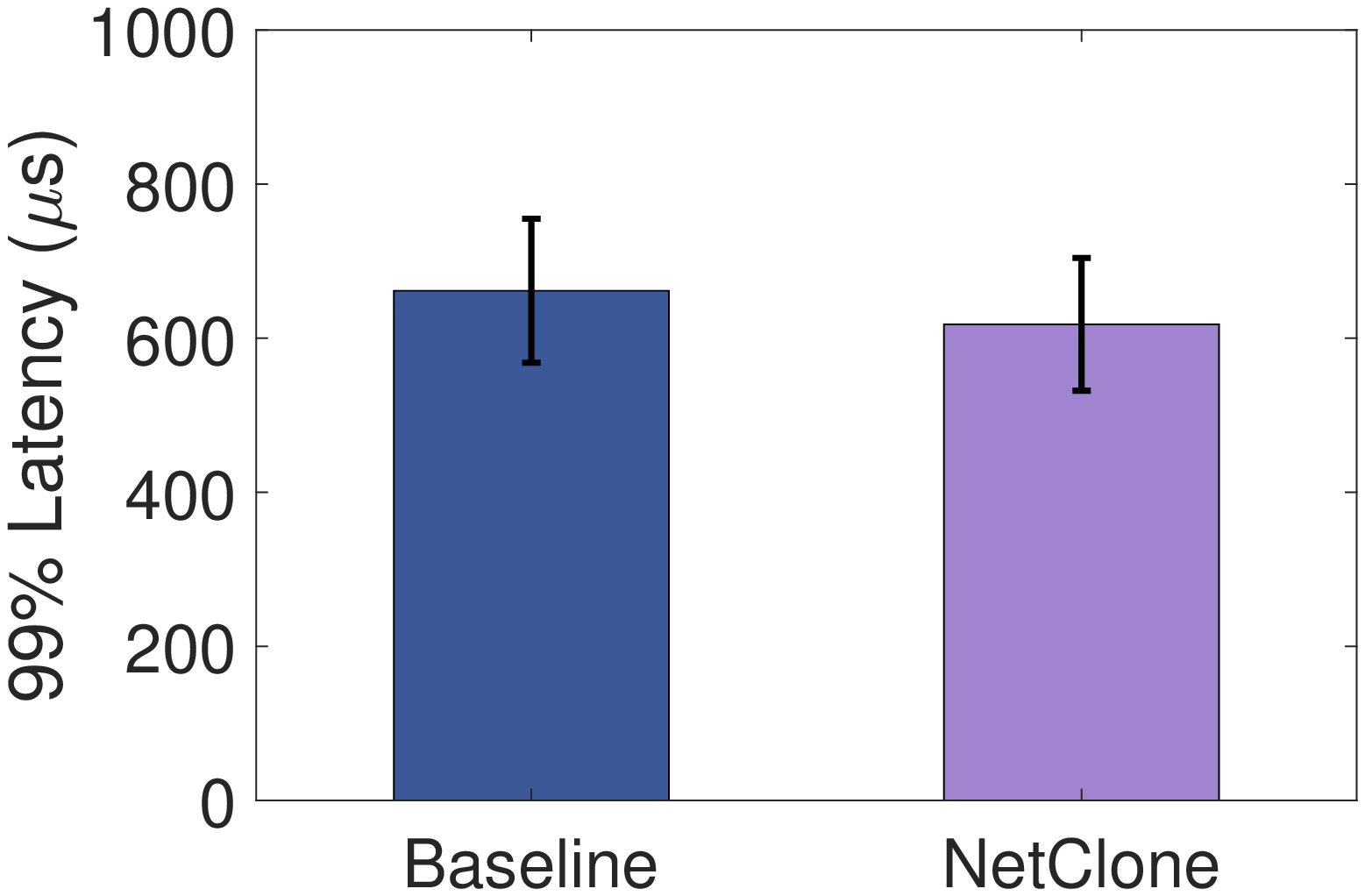}}\hfill
\caption{Confidence of the empty queue for state signaling.
\label{fig:qlen}}
\end{figure}

\begin{figure}[t]
\centering
\subfloat[Exp(25)]{\includegraphics[width=0.500\linewidth]{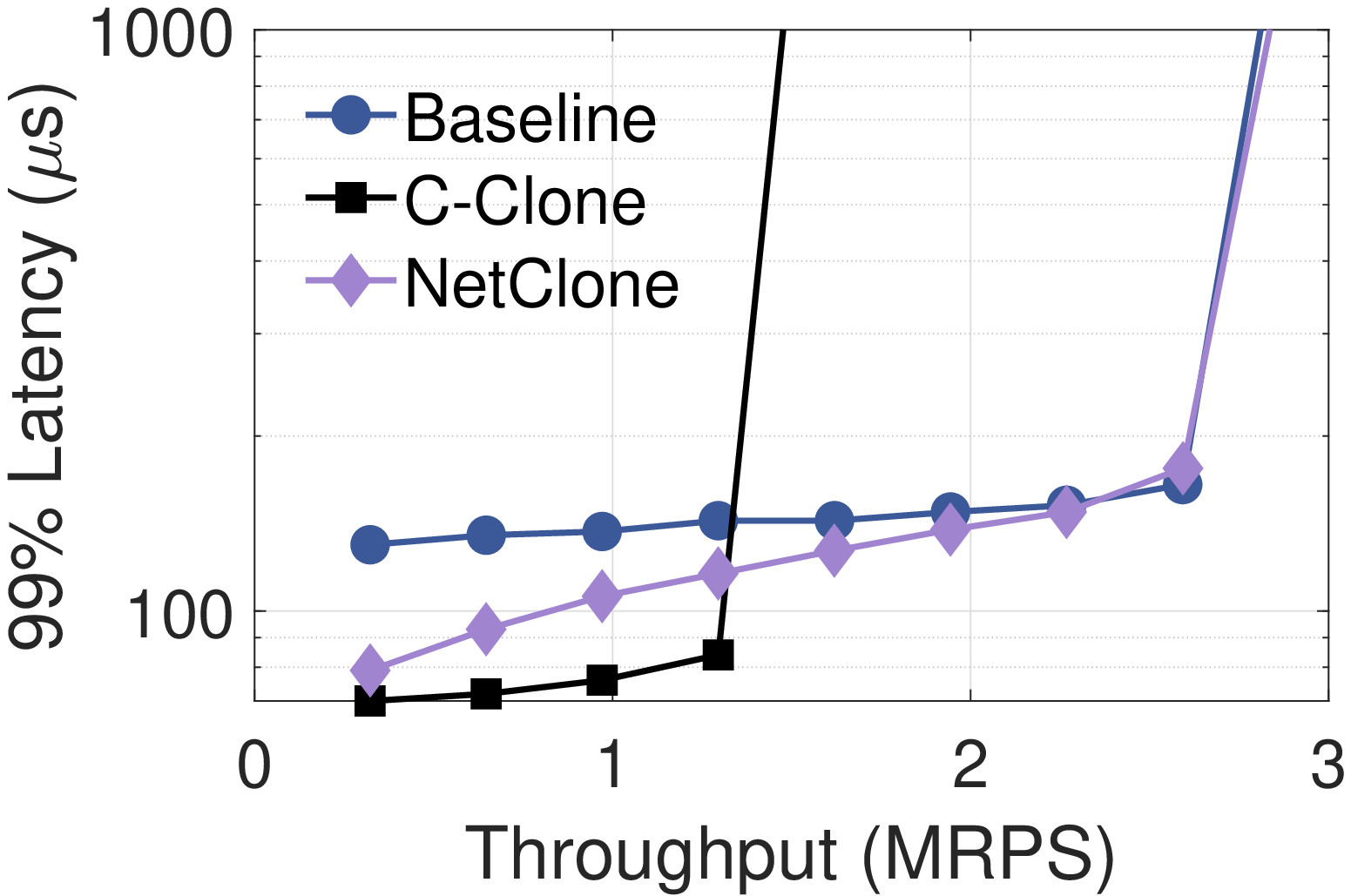}}\hfill
\subfloat[Bimodal(90\%-25,10\%-250)]{\includegraphics[width=0.500\linewidth]{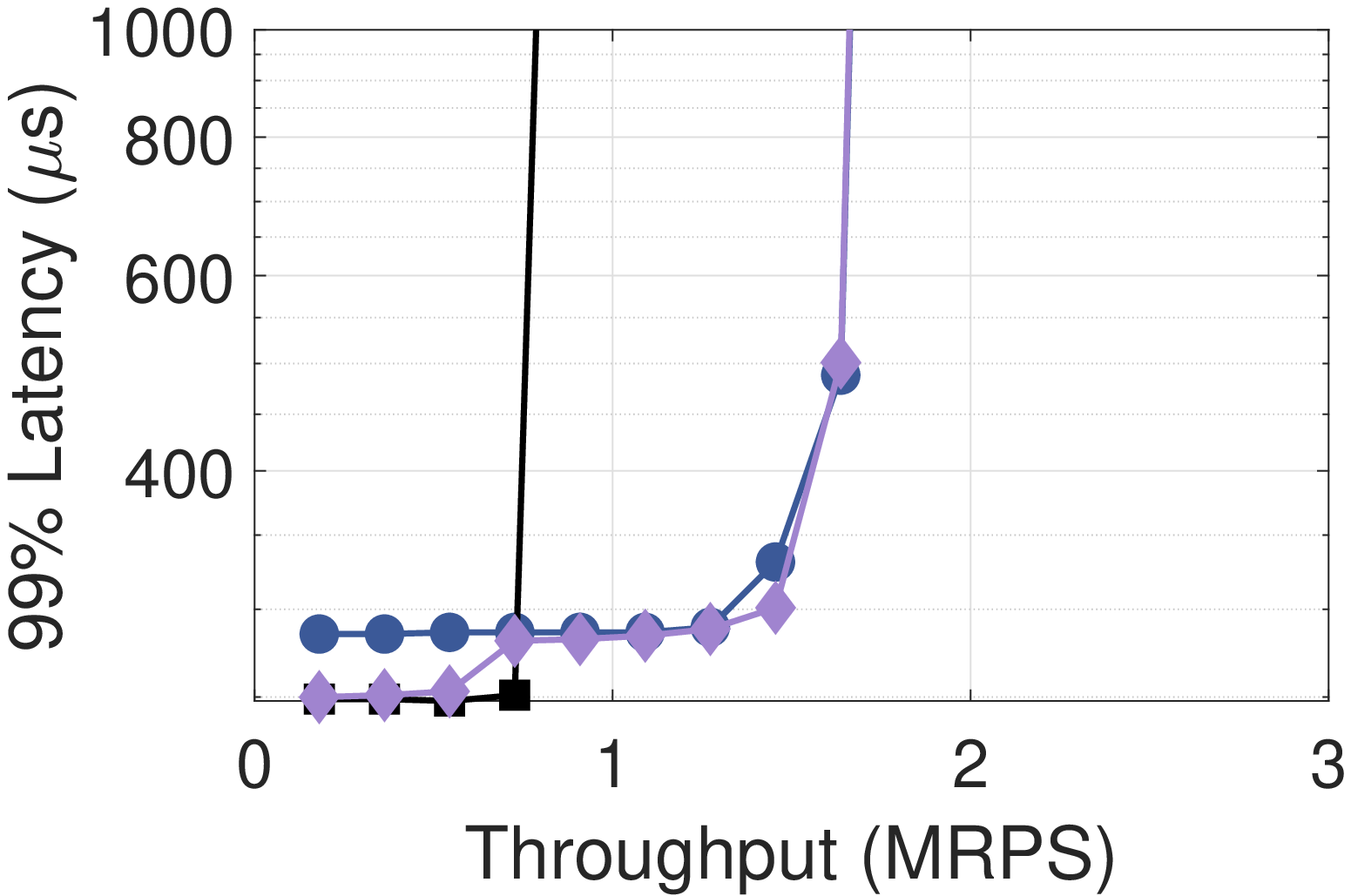}}\hfill
\caption{Experimental results with a low service-time variability ($p$=0.001).
\label{fig:main-001}}
\end{figure}

\subsection{Deep Dive}
\subsubsection{Confidence of State Signals}
\sys considers the server as idle if the queue length of the server is zero.
Therefore, we investigate the portion of empty queues by varying loads.
We make a server record its current queue length when sending a response.
In Figure~\ref{fig:qlen} (a), we can see that the portion of empty queues decreases as the load grows, as expected.
We see two important observations as follows.
First, even at low loads, the queue may not be empty.
This explains why \sys shows higher latency than \cc at low loads in Figure~\ref{fig:main}.
Second, likewise, queues do not always build up even under very high loads.
This is the reason why cloning happens at not only low loads but also high loads.
To check the efficiency of cloning at high loads, we run experiments with the baseline and \sys 10 times at 0.9 of load and get the average tail latency and their standard deviations.
Figure~\ref{fig:qlen} (b) shows the results.
As expected, we see that \sys may cause worse latency than the baseline occasionally.
However, by considering the average and the standard deviation, we can conclude that \sys generally provides better latency than the baseline even at very high loads.

\subsubsection{Impact of Service-Time Variability}
Figure~\ref{fig:main-001} shows the experimental results for synthetic workloads with a low variability of $p=0.001$.
The Y-axis of Figure~\ref{fig:main-001} (a) and (b) is in the log-scale.
We can see that \sys can decrease tail latency even if the service-time variability is low.
The trend of experimental results is similar to Figure~\ref{fig:main}.
One difference is that performance improvement slightly decreases.
However, it is not surprising since the benefit of request cloning comes from masking service-time variability.

\begin{figure}[t!]
\centering
\includegraphics[width=8.0cm]{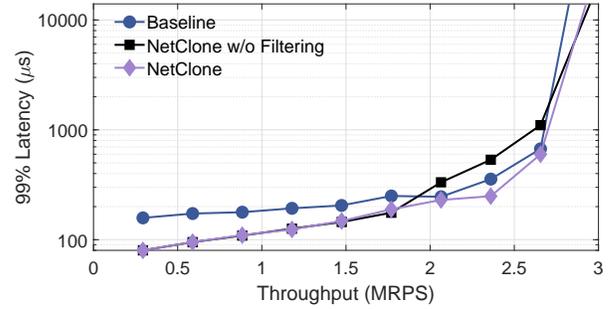}
\caption{Impact of redundant response filtering. \label{fig:overhead}}
\end{figure}

\begin{figure}[t!]
\centering
\includegraphics[width=8.0cm]{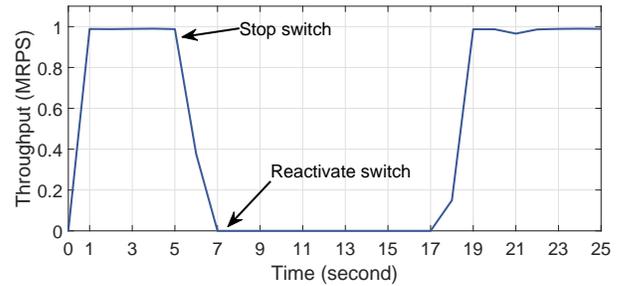}
\caption{Performance under switch failures. \label{fig:failure}}
\end{figure}

\subsubsection{Impact of Redundant Response Filtering}
We now inspect the impact of redundant response filtering.
To do this, we turn off the response filtering function and compare its performance against the baseline and \sys.
Figure~\ref{fig:overhead} plots the result.
We have the following observations.
First, at low loads, redundant responses barely harm performance since the client has enough capability to handle redundancy.
However, as the system load grows, the latency gets worse.
The performance is even worse than the baseline at high loads if \sys does not use response filtering.
This means that filtering redundant responses plays an important role to optimize the performance of \sys.

\subsubsection{Performance under Switch Failures}
In this experiment, we evaluate the resilience of \sys to switch failures.
Figure~\ref{fig:failure} shows the throughput for 25 seconds.
The switch was stopped at 5 seconds and manually reactivated at 7 seconds.
The throughput recovers after approximately 10 seconds.
The downtime is not a result of \sys, but rather depends on the switch architecture.
Thus, we can say that \sys is robust to switch failures.
Note that \sys does not incur permanent misbehavior since the switch stores only soft states.

%% file: sections/relatedwork.tex
\section{Related work\label{relatedwork}}
We briefly discuss existing works related to \sys in terms of request cloning, server-level solutions, and in-network computing solutions.

\textbf{Request cloning.}
Vulimiri \etal~\cite{vulimiri13} investigates the tradeoff of client-based request cloning.
They identify the threshold load and the client-side overhead.
Gardner \etal~\cite{powerofd,redundancy} provide rigorous theoretical analysis for cloning.
Dolly~\cite{dolly} and RepFlow~\cite{xu14a} utilize the cloning technique for mitigating stragglers in MapReduce clusters and multi-path routing in data center networks, respectively.
\sota~\cite{laedge} performs dynamic cloning using the coordinator but lacks low latency overhead and scalability.
\sys is the first dynamic request cloning system for microsecond-scale RPCs.

\textbf{Server-level solutions for microsecond-scale RPCs.}
There are line of works that reduce the latency of RPCs at the server level in hardware and software.
ALTOCUMULUS~\cite{alto} avoids the scheduling overhead using direct register-level messaging.
RPCValet~\cite{rpcvalet} bypasses slow PCIe buses using shared caches when dispatching RPCs to CPU cores.
nanoPU~\cite{nanopu} bypasses the cache and memory hierarchy to provide a fast path from NIC to applications using a hardware accelerator.
eRPC~\cite{erpc} improves the performance of small messages by optimizing common cases with various software techniques.
IX~\cite{ix}, ZygOS~\cite{zygos}, and Shinjuku~\cite{shinjuku} are data plane OSes that provide efficient CPU scheduling for microsecond-scale RPCs.
For example, Shinjuku~\cite{shinjuku} implements a preemptive scheduling algorithm by re-queueing long-lasting RPCs if the runtime exceeds a given threshold.
The above works address the RPC latency at the server level, whereas \sys tries to optimize RPC latency at the cluster level.
Since \sys does not restrict the server-side mechanism to a specific solution, \sys is orthogonal to the existing works.

\textbf{In-network computing for microsecond-scale RPCs.}
The capability and flexibility of programmable switch ASICs trigger the emergence of in-network computing.
NetCache~\cite{jin17}, Pegasus~\cite{pegasus}, DistCache~\cite{liu19}, Harmonia~\cite{harmonia}, NetLR~\cite{netlr}, P4DB~\cite{p4db}, and Transaction Triaging~\cite{triaging} are solutions to accelerate distributed storage.
\sys can improve the latency of GET queries and is a more generic solution.
RackSched~\cite{racksched} is an in-network request scheduler for microsecond-scale RPCs that performs the JSQ load balancing.
\sys is orthogonal to RackSched since \sys does not specify its load balancing algorithm.

%% file: sections/conclusion.tex
\section{Conclusion\label{conclusion}}
In this paper, we presented \sys, a new request cloning system that dynamically and quickly replicates requests to reduce the tail latency of microsecond-scale RPCs at scale.
Unlike traditional client-based or coordinator-based cloning approaches, \sys performs request cloning in the network switch using programmable switch ASICs.
Various technical challenges to design and implement \sys in the switch data plane were addressed.
We have implemented a prototype of \sys with an Intel Tofino switch and a cluster of commodity servers.
The experimental results showed that \sys effectively improves the tail latency of RPCs for both synthetic and real-world application workloads.
We believe that network switches would play a key role as domain-specific hardware for microsecond-scale RPCs.
We emphasize that there are remaining problems to realize the vision of in-network computing, which include fully synthesizing existing in-network solutions for microsecond-scale RPCs and integrating \sys with existing RPC frameworks.
\\
\textbf{Ethics:} This work does not raise any ethical issues.

%% file: sections/ack.tex
\section*{Acknowledgement}
We would like to thank our shepherd, Andreas Haeberlen, and the anonymous SIGCOMM reviewers for their insightful comments and constructive feedback.
This research was sponsored by the National Research Foundation of Korea (NRF) grants funded by the Ministry of Science and ICT (No. RS-2023-00240029).
Gyuyeong Kim is the corresponding author.